\documentclass[smallextended]{svjour3}       
\smartqed  
\usepackage{graphicx}

 \usepackage{mathptmx}      
  
\usepackage{latexsym}  
  
\usepackage{amsmath}  
\usepackage{amssymb}  
\usepackage{amscd}  
\newcommand{\1}{\mbox{1}\hspace{-0.25em}\mbox{l}}  
 \journalname{Acta Applicandae Mathematica}  
\begin{document}  
\title{Local description of band rearrangements}  
  
  

\subtitle{ Comparison of  
semi-quantum and full quantum approach.}  
\titlerunning{Band rearrangements.}  
\author{Toshihiro Iwai        \and  
        Boris Zhilinskii 
}  
  
\institute{T. Iwai \at  
              Kyoto University, 606-8501 Kyoto, Japan \\  
              \email{iwai.toshihiro.63u@st.kyoto-u.ac.jp}       
              \and  
           B. Zhilinkii \at  
              Universit\'e du Littoral C\^ote d'Opale,  
              189A av. M. Schumann, Dunkerque 59140 France \\  
              Tel.: 33-3-28658266\\  
              Fax.: 33-3-28658244\\  
              \email{zhilin@univ-littoral.fr}   
}  
  
\date{Received: date \today / Accepted: date}  

\maketitle  
  
\begin{abstract}  
Rearrangement of rotation-vibration energy bands in isolated  
molecules within semi-quantum approach is  
characterised by delta-Chern invariants associated to a   
local semi-quantum Hamiltonian valid in a small neighborhood  
of a degeneracy point for the initial semi-quantum Hamiltonian and also valid in a small neighborhood   
of a critical point corresponding to the crossing of the boundary between iso-Chern domains   
in the control parameter space.  For a full quantum model,   
a locally approximated Hamiltonian is assumed to take the form of a Dirac operator   
together with a specific boundary condition.  It is demonstrated that the crossing of   
the boundary along a path with a delta-Chern invariant equal to $\pm1$ corresponds  
to the transfer of one quantum level from a subspaces of quantum states to the other subspace   
associated with respective positive and negative energy eigenvalues of the local Dirac Hamiltonian.   
   
\keywords{Energy band \and Chern number \and Dirac operator}  
 \PACS{03.65.Aa \and 03.65.Vf \and 33.15.Mt}  
\subclass{53C80 \and 81Q70 \and 81V55}  
\end{abstract}  
  
\section{Introduction}  
\label{intro}  
Redistribution of energy levels between energy bands in rotation-vibration  
structure of isolated molecules under the variation of a control parameter  
was a subject of a number of publications   
\cite{VPVdp,FaurePRL,IwaiAnnPhys,IZ2013,IwaiTCA2014}.   
In semi quantum models \cite{PhysRev93,SadZhilMonodr,PhysRep2},   
rotational and vibrational variables are treated as classical and quantum ones,   
respectively, and the semi-quantum Hamiltonian takes the form of a Hermitian matrix.     
With each non-degenerate eigenvalue of that Hamiltonian,    
there are associated an eigen-line bundle, whose base space is a classical phase space   
for rotational variables and fibers represent vibrational quantum states   
\cite{FaurePRL,IwaiAnnPhys}.  The Chern numbers of respective eigen-line bundles   
are topological invariants characterizing the band structure.   
  
The qualitative rearrangement of energy bands is associated with the formation of degeneracy  
points of eigenvalues of the matrix Hamiltonian.   
The critical points in the control parameter space, for which the Hamiltonian has   
degeneracy points on the classical phase space, form boundaries between iso-Chern domains.   
For a one-parameter family of Hamiltonians attached with a path in the control parameter space,   
the crossing of the boundary between iso-Chern domains can be characterized   
by a local delta-Chern, which is a topological invariant associated with a locally approximated Hamiltonian   
defined  in a small  neighborhood of a degeneracy point for the Hamiltonian and also defined   
in a small neighborhood of the crossing point at the boundary.     
The totality of the local delta-Cherns provides the modification of the Chern number of the   
eigen-line bundle in question during the rearrangement \cite{IZ2013},   
if the symmetry group action is taken into account in addition.   
  
The present paper introduces on a simple generic example a full quantum local description  
of the rearrangement phenomenon by constructing a local Dirac operator in   
association with doubly degenerate eigenvalues of the semi-quantum local Hamiltonian.    
On introducing appropriate boundary conditions, the delta-Chern invariant calculated for   
the corresponding semi-quantum local Hamiltonian is shown to be exactly the same as the spectral   
flow for the local Dirac operator, {\it i.e.},  the difference between the number of positive and negative  
eigenvalues when the control parameter passes zero along the path crossing the boundary.    
  
An initial effective quantum Hamiltonian describing a rotation structure of   
several vibrational states is generally put in a matrix form.   
For two quantum vibrational states,  a generic Hamiltonian takes the form   
\begin{equation} \label{twoStateH}  
     H(t;J_\alpha)=\left(\begin{array}{cc} h_{11}(t;J_\alpha) & h_{12}(t;J_\alpha)  
               \\  h_{21}(t;J_\alpha) & - h_{11}(t;J_\alpha) \end{array}  
    \right),   
\end{equation}  
where the matrix elements are rotational operators. The whole  
matrix elements are determined by specifying the representations of the symmetry group   
on vibrational and rotational variables.    
A semi-quantum Hamiltonian follows from an effective quantum one   
by replacing the rotational quantum operators by classical variables defined on classical phase space   
for rotational motions, which is a two-dimensional sphere $S^2$. Generically, isolated degeneracy points   
appear  somewhere on the sphere $S^2$ at an isolated value of the critical parameter $t$   
which is taken as $t=0$.     
This fact is a consequence of the fact that the codimension of degeneracy of  
eigenvalues of an Hermitean matrix is three \cite{ArnoldChern}.   
A family of the simplest quantum Hamiltonians possessing in the semi-quantum limit  
an isolated degeneracy point takes the form  
 \begin{equation}  
     H(t,J_{\alpha})=\left(\begin{array}{cc}  t+J_z-J & J_x+iJ_y \\ J_x-iJ_y & -t-J_z-J \end{array}  
    \right).    
\end{equation}  
This one-parameter family of quantum Hamiltonians demonstrates a redistribution  
of one energy level between two bands \cite{VPVdp} and its semi-quantum analog  
is characterized by a modification of Chern numbers associated with each  
band \cite{FaurePRL,FaureCP2,IwaiAnnPhys}.    
This model has a tight mathematical relation to a description of topological phase transitions   
in solid state physics \cite{IwaiTCA2014}, in particular to   
quantum Hall effect, topological insulators, {\it et al.}   
The last but not least aim of the present paper is to explore this mathematical relationship   
by describing a generic phenomenon of energy band rearrangement through the study of a   
Dirac-type Hamiltonian as a full quantum local model associated with a linearization of   
the generic Hamiltonian (\ref{twoStateH}).    
  
The organization of this article is as follows:   
A brief review is made of the delta-Chern for a semi-quantum Hamiltonian in Sec.~\ref{review_delta_chern},   
and a Dirac Hamiltonian as an associated local full quantum Hamiltonian is introduced in   
Sec.~\ref{full-quantum Ham}.   
After discussing the $SO(2)$ symmetry of the Dirac operator in Sec.~\ref{SO(2) symmetry},   
the separation of variables method is applied in Sec.~\ref{separation of variables}   
to solve the Dirac equation.   
Section \ref{boundary conditions} is devoted to a search for boundary conditions for   
the Dirac equation defined on a bounded domain and an APS boundary condition is found.   
In Secs.~\ref{edge states}, \ref{zero modes}, and \ref{regular states},   
edge states, zero modes, and regular states are worked out, respectively.   
The behavior of eigenvalues as functions of the control parameter is summarized   
in Sec.~\ref{spectral flow}.   Section \ref{conclusion} includes concluding remarks and the comparison   
of the band rearrangement with the topological insulators.

\section{A brief review of delta-Chern for a semi-quantum Hamiltonian}  
\label{review_delta_chern}  
We here consider a simple model Hamiltonian which is supposed to be the linear approximation   
of an original semi-quantum model Hamiltonian, which results from (\ref{twoStateH}),    
at a degeneracy point on the two-sphere;   
\begin{equation}  
\label{semi-quantum Ham}  
     H(t,p)=\left(\begin{array}{cc}  t & p_1-ip_2 \\ p_1+ip_2 & -t \end{array}  
    \right),   
\end{equation}  
where $(p_1,p_2)$ denotes the Cartesian coordinates on the tangent plane at the   
degeneracy point in question, and $t$ is a parameter.    
The Hamiltonian of this type dates back to \cite{Berry,GeometricPhase} on geometric phases.   
As is easily seen, the eigenvalues of $H$ are given by   
\begin{equation}  
\label{semi-quantum energy}  
    E^{\pm}=\pm \sqrt{t^2+|p|^2}, \quad |p|^2=p_1^2+p_2^2.   
\end{equation}  
The degeneracy in eigenvalues occurs when $t=0$ at $(p_1,p_2)=(0,0)$ only.   
  
The ``up" eigenvectors (we follow here the terminology and conventions  
introduced in \cite{IwaiAnnPhys})   
associated with the positive and the negative eigenvalues $E^{\pm}$   
are expressed as   
\begin{equation}  
 |u^{\pm}_{\rm up}(t,p)\rangle =\frac{1}{N_{\rm up}^{\pm}}\begin{pmatrix}   
   p_1-ip_2 \\ E^{\pm}-t \end{pmatrix}, \quad   
   N_{\rm up}^{\pm}=\sqrt{ |p|^2 +(E^{\pm}-t)^2}.      
\end{equation}  
We choose the eigen-line bundle associated with $E^+$ for the   
calculation of the local contribution to the Chern number.       
The exceptional point of  $|u^+_{\rm up}(t,p)\rangle$ exists at $(p_1,p_2)=(0,0)$ for $t>0$ only,   
which means that the local Chern number is zero for $t<0$.    
Since the orientation of a small circle centered at the exceptional point is clockwise,   
the winding number associated with the exceptional point is $+1$ for $t>0$, so that the   
(local) Chern number is $-1$. When the parameter $t$ passes the degeneracy point $t=0$ of the   
control parameter from the negative side $t<0$ to the positive side $t>0$, we have the (local)   
delta-Chern  $-1-0=-1$.   
  
For comparison sake, we treat the same problem by using the ``down" eigenvectors.     
The ``down" eigenvectors associated with the positive and the negative eigenvalues $E^{\pm}$   
are expressed as   
\begin{equation}  
   |u^{\pm}_{\rm down}(t,p)\rangle =\frac{1}{N^{\pm}_{\rm down}} \begin{pmatrix}   
    E^{\pm} +t  \\ p_1+ip_2 \end{pmatrix}, \quad   
   N_{\rm down}^{\pm} =\sqrt{ |p|^2 + (E^{\pm}+t)^2}.      
\end{equation}     
The exceptional point of  $|u^+_{\rm down}(t,p)\rangle$ appears at $(p_1,p_2)=(0,0)$ for $t<0$ only,   
which means that the local Chern number for $t>0$ is zero.     
Since the orientation of a small circle centered at the exceptional point   
is clockwise,   
the winding number associated with the exceptional point is $-1$ for $t<0$, so that the   
(local) Chern number is $+1$ for $t<0$.  When the parameter $t$ goes through the degeneracy point $t=0$   
from the negative side $t<0$ to the positive side $t>0$, the accompanying local delta-Chern is   
$0-1=-1$.    
Thus, the delta-Chern $-1$ is assigned to the eigen-line bundle associated with the positive eigenvalue $E^+$.   
  
Needless to say, the delta-Chern $+1$ is assigned to the eigen-line bundle associated   
with the negative eigenvalue $E^-$.     
  
Summing up the above discussion, we have the following table for the winding numbers;  
\begin{equation}   
  \begin{array}{c|c|c|c||c}  
         &   t<0  & t=0   & t>0 & \\ \hline   
         &    W^+_{\rm up}=0 & {\rm no} & W^+_{\rm up}=1 & \Delta W^+_{\rm up}=1-0=1 \\ \cline{2-5}   
      {\vspace{-5mm} E^+}  &   W^+_{\rm down}=-1 & {\rm no} & W^+_{\rm down}=0 &   
     \Delta W^+_{\rm down}=0-(-1)=1 \\ \hline  
        &    W^-_{\rm up}=1 & {\rm no} & W^-_{\rm up}=0 &  \Delta W^-_{\rm up}=0-1=-1\\ \cline{2-5}   
       {\vspace{-5mm} E^-}  &   W^-_{\rm down}=0 & {\rm no} & W^-_{\rm down}=-1 &   
       \Delta W^-_{\rm down}=-1-0=-1     \\  \hline  
  \end{array}  
\end{equation}  
\medskip  
\par\noindent  
The delta-Chern is of course defined to be $\Delta c^{\pm} =-\Delta W^{\pm}_{{\rm up}/{\rm down}}$,   
where the superscripts $\pm$ indicate that the quantities in question   
are assigned to the energy eigenvalues $E^{\pm}$, respectively.       
  
\begin{figure}[htbp]  
\begin{center}  
\includegraphics[width=0.8\columnwidth]{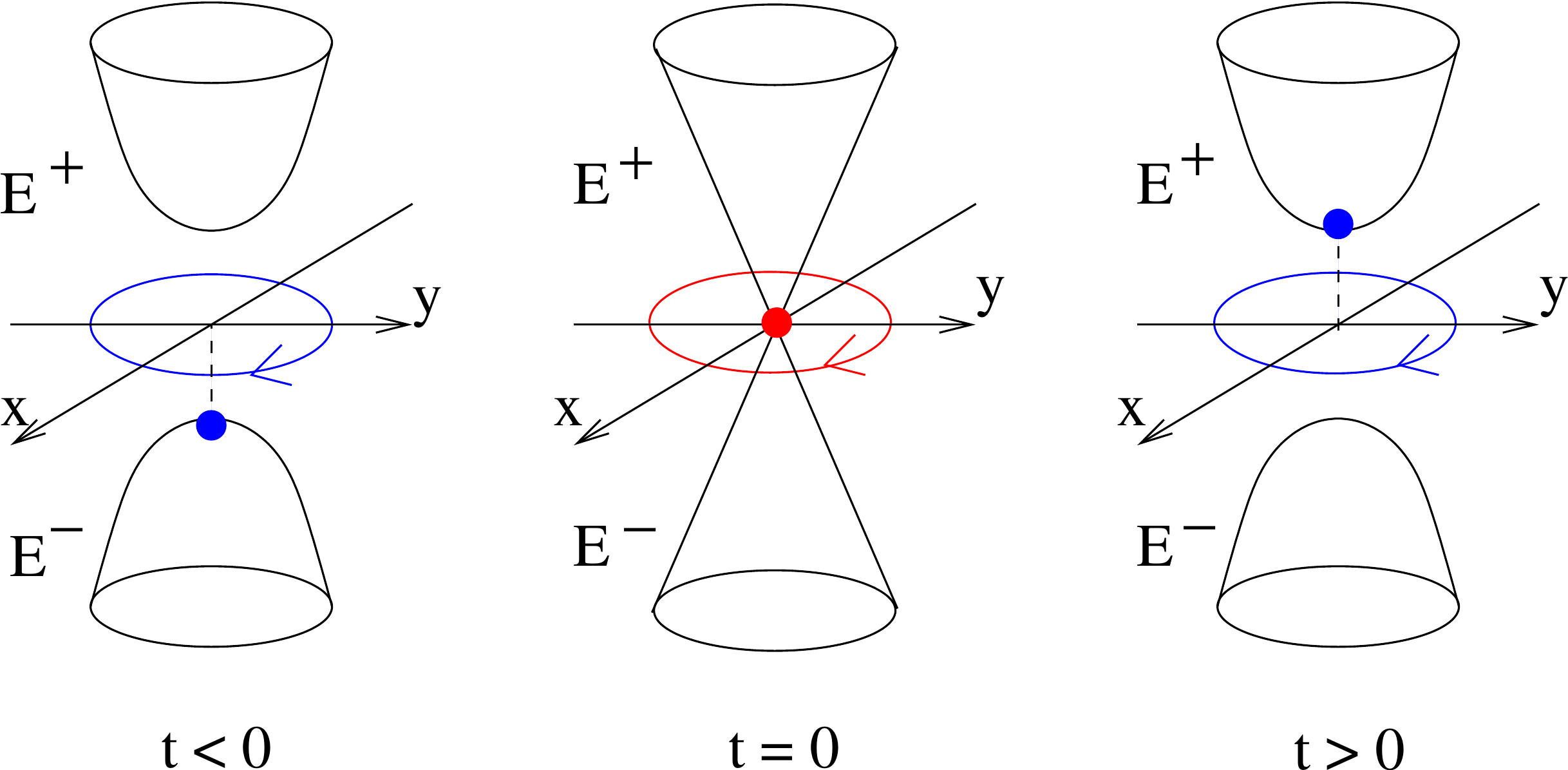}  
\end{center}  
\caption{A schematic description of the transition of the exceptional point for the ``up"   
eigenvector against $t$.  
The small circle is clockwise oriented for the ``up" eigenvector.}  
\label{excep pts}  
\end{figure}

Since the exceptional point is responsible for the local delta-Chern,   
we are interested in which of the eigenvectors $|u^{\pm}_{\rm up}(t,p)\rangle$   
the exceptional point is attached to, according to the variation of the control parameter.     
From this point of view, we obtain the table;   
\begin{equation}  
\label{ex pt shift}  
  \begin{array}{c|c|c|c}  
         & t<0  & t=0 & t>0 \\ \hline   
    \mbox{ex. pt. for}\; |u^{\pm}_{\rm up}\rangle & - & \mbox{deg. pt.} & +  \\ \hline   
    \mbox{ex. pt. for}\; |u^{\pm}_{\rm down}\rangle & + & \mbox{deg. pt.}  & -  \\ \hline    
  \end{array}  
\end{equation}  
This table shows, for example, that as for the ``up" eigenvector the exceptional point   
assigned to the eigenvector $|u^{-}_{\rm up}(t,p)\rangle$ associated with the negative eigenvalue   
$E^-$ for $t<0$ shifts to that assigned to $|u^{+}_{\rm up}(t,p)\rangle$ associated with   
$E^+$ for $t>0$ (see Fig.~\ref{excep pts}).    
As for the ``down" eigenvector, a similar shift of the exceptional  point occurs in the opposite direction.    
  
We note in addition that the calculation of the winding number assigned to each exceptional  
point of the eigenvector concerned is a frequently used method to find the Chern number,  
which is effectively employed for (global) semi-quantum models   
in the presence of cubic symmetry  \cite{Iwai}.

\section{ An associated full quantum Hamiltonian}  
\label{full-quantum Ham}  
\par  
We are interested in the full quantum Hamiltonian associated with the semi-quantum  
local Hamiltonian \eqref{semi-quantum Ham}.      
Let us be reminded of the fact that the original semi-quantum Hamiltonian is obtained by replacing   
the operators $J_k$ by the classical variables $p_k,\,k=1,2,3$, approximately speaking.    
While the linear local Hamiltonian $H(t,p)$ should be viewed as a semi-quantum Hamiltonian,   
the correspondence between operators and the classical variables $p_k$ is not well understood   
on the level of linear approximation.    
In order to re-quantize the $H(t,p)$, we have to assume some correspondence rule between classical   
variables and quantum operators. To this end, we recall that   
the two-sphere $S^2$ is viewed as a (co)adjoint orbit of the rotation group $SO(3)$ or   
as $SO(3)$ coherent states.     
A question arises as to whether the tangent plane at the degeneracy point is viewed   
as an orbit of some symmetry group.    
The Euclidean motion group $E(2)\cong SO(2)\ltimes \mathbb{R}^2$ would be a candidate of   
such a symmetry group.    
If we assume that the total group $SO(3)$ turns into $E(2)$ in the linear approximation   
at the degeneracy point, the tangent plane can be viewed as an orbit of $E(2)$.   
Then, the correspondence between the classical variables $p_k$ and suitable quantum operators should   
be considered within generators of $E(2)$.   
We then come to the quantization idea that the variables $p_k$ could be suitably replaced by   
the generators, $p_k=-i\partial /\partial q_k$, $k=1,2$, of $\mathbb{R}^2$.   
Then, the full quantum Hamiltonian  associated with $H(t,p)$ is expressed as   
\begin{equation}  
\label{full-q-Ham}  
    \hat{H}_t=\left(\begin{matrix}  t   
   & -i\frac{\partial}{\partial q_1}- \frac{\partial}{\partial q_2} \\  
     -i\frac{\partial}{\partial q_1} + \frac{\partial}{\partial q_2}   
         & -t \end{matrix}\right) .    
\end{equation}  
In terms of the Pauli matrices,   
$$   \sigma_1=\left(\begin{matrix} 0 & 1 \\ 1 & 0 \end{matrix}\right), \quad   
    \sigma_2=\left(\begin{matrix} 0 & -i \\ i & 0 \end{matrix}\right), \quad   
    \sigma_3=\left(\begin{matrix} 1 & 0 \\ 0 & -1 \end{matrix}\right), $$  
the Hamiltonian $\hat{H}_t$ is put in the form of (time-independent) Dirac Hamiltonian   
in two space-dimensions   
\begin{equation}   
    \hat{H}_t =-i\sum_{j=1}^2 \sigma_j \frac{\partial}{\partial q_j} +t \sigma_3,    
\end{equation}  
where the parameter $t$ plays the role of (variable) mass.    
Since the operator $\hat{H}_t$ is considered as a linear approximation, it may   
be accompanied by a suitable boundary condition, which will be discussed later.   
   
By introducing the polar coordinates $(r,\theta)$, the Hamiltonian $\hat{H}_t$ is rewritten as     
\begin{equation}   
    \hat{H}_t = \begin{pmatrix}   
    t & e^{-i\theta}\bigl(-i\frac{\partial}{\partial r}-\frac{1}{r}\frac{\partial}{\partial \theta}\bigr) \\   
   e^{i\theta}\bigl(-i\frac{\partial}{\partial r}+\frac{1}{r}\frac{\partial}{\partial \theta}\bigr) & -t   
  \end{pmatrix}.   
\end{equation}  
To put the Hamiltonian in a manifest manner, it is convenient to introduce   
the Pauli matrices associated with the polar coordinates,   
\begin{equation}   
    \sigma_r=\vec{n}\cdot \vec{\sigma}=  
  \begin{pmatrix} 0 & e^{-i\theta} \\ e^{i\theta} & 0 \end{pmatrix},  \quad   
    \sigma_{\theta}=\vec{t}\cdot \vec{\sigma}=  
  \begin{pmatrix} 0 & -i e^{-i\theta} \\ i e^{i\theta} & 0 \end{pmatrix},   
\end{equation}   
where $\vec{n}$ and $\vec{t}$ are given by   
\begin{equation}   
      \vec{n} = \frac{1}{r}\begin{pmatrix} q_1 \\ q_2 \end{pmatrix}, \quad   
      \vec{t} = \frac{1}{r}\begin{pmatrix} -q_2 \\ q_1 \end{pmatrix},   
\end{equation}  
which are the unit normal vector to the circle of radius $r=\sqrt{q_1^2+q_2^2}$ centered at the origin   
and the unit tangent vector to the same circle, respectively,   
and the frame $\{\vec{n},\vec{t}\}$ is positively oriented; $\det(\vec{n},\vec{t})=1$.     
In terms of the new sigma matrices, the Hamiltonian is expressed as   
\begin{equation}  
\label{polar-Ham}   
   \hat{H}_t=   
    -i\sigma_r\frac{\partial}{\partial r} -\frac{i}{r}\sigma_{\theta}\frac{\partial}{\partial \theta}  
    +t \sigma_3.  
\end{equation}  
  
\section{$SO(2)$ symmetry of the semi-quantum and the full quantum Hamiltonians}  
\label{SO(2) symmetry}  
\par  
As is easily verified, the semi-quantum Hamiltonian $H(t,p)$ is $SO(2)$-invariant in the sense that   
\begin{equation}   
   D(e^{i\tau})H(t,p)D(e^{i\tau})^{-1}=H(t,R(\tau)p),   
\end{equation}  
where   
\begin{equation}  
  D(e^{i\tau}):=e^{-i\tau\sigma_3 /2}=\begin{pmatrix} e^{-i\tau/2} & 0 \\ 0 & e^{i\tau/2} \end{pmatrix}, \quad   
  R(\tau):=\begin{pmatrix}  \cos \tau & -\sin \tau \\ \sin \tau & \cos \tau \end{pmatrix}, \quad   
  \tau\in \mathbb{R}.   
\end{equation}  
We turn to the $SO(2)$ symmetry of the full quantum Hamiltonian $\hat{H}_t$.   
Let $\Phi$ denote a two-component spinor defined on $\mathbb{R}^2$.   
The $U(1)$ action on the spinor $\Phi$ is defined through the diagram   
\begin{equation}   
 \begin{CD}  
      \mathbb{R}^2 @>\Phi >> \mathbb{C}^2 \\  
       @VR(\tau)VV   @VV{D(e^{i\tau})}V \\  
     \mathbb{R}^2 @>>U_{\tau}\Phi> \mathbb{C}^2  
 \end{CD}\qquad .  
\end{equation}  
In other words, the $U(1)$ action on the spinor is defined to be   
\begin{equation}   
     U_{\tau}\Phi=D(e^{i\tau})\Phi\circ R(-\tau).    
\end{equation}   
As is straightforwardly verified, the $U(1)$ symmetry of the $\hat{H}_t$ is described as   
\begin{equation}  
\label{sym-H}  
    U_{\tau}\hat{H}_tU_{\tau}^{-1}=\hat{H}_t.   
\end{equation}  
The infinitesimal generator $J$ of $U_{\tau}$ is called the (spin-orbital) angular momentum operator,   
which is determined by   
$U_{\tau}=\exp(-i\tau J)$.  Since   
\begin{equation}    
   \frac{d}{d\tau}U_{\tau}\Bigr|_{\tau=0}=-\frac{i}{2}\sigma_3+   
     \1 \Bigl(q_2\frac{\partial}{\partial q_1}-q_1\frac{\partial}{\partial q_2}\Bigr),   
\end{equation}  
we obtain   
\begin{equation}   
    J=\frac12\sigma_3 +i \1 \Bigl(q_2\frac{\partial}{\partial q_1}-q_1\frac{\partial}{\partial q_2}\Bigr)  
    =\frac12\sigma_3 -i \1 \frac{\partial}{\partial \theta}.   
\end{equation}  
The differentiation of  \eqref{sym-H} with respect to $\tau$ at $\tau=0$ yields   
\begin{equation}   
         [J,\hat{H}_t]=0.   
\end{equation}

\section{Separation of variables method}  
\label{separation of variables}  
\par  
Since $[J,\hat{H}_t]=0$, the eigenvalue problem $\hat{H}_t\Phi=E\Phi$ can be reduced to the subproblems   
on the eigenspaces of $J$.   
Let us denote the eigenvalue of $J$ by $j$.  Then, the eigenvalue equation $J\Phi=j\Phi$ is solved by   
\begin{equation}  
\label{eigen-st(j)}   
      \Phi_j(r,\theta)=  
  \begin{pmatrix}  e^{i(j-\frac12)\theta} \phi^{(-)}_j(r) \\  e^{i(j+\frac12)\theta} \phi^{(+)}_j(r) \end{pmatrix}.   
\end{equation}  
Since $(r,\theta)$ and $(r,\theta+2n\pi)$ with $n\in \mathbb{Z}$ are the same point of the plane   
$\mathbb{R}^2$, the eigenvalue $j$ should be a non-integer half-integer;   
$j\in \{\pm\frac12,\pm\frac32, \cdots\}$.  
  
Thus, the initial equation $\hat{H}_t\Phi=E\Phi$ reduces to the coupled equations for $\phi_j^{\pm}$   
\begin{subequations}  
\label{Eq-phis}  
\begin{align}  
    -i\frac{d\phi^{(+)}_j}{dr}-\frac{i}{r}(j+\frac12)\phi^{(+)}_j+t\phi_j^{(-)}=E_j\phi_j^{(-)}, \label{Eq-phi+}\\  
    -i\frac{d\phi^{(-)}_j}{dr}+\frac{i}{r}(j-\frac12)\phi^{(-)}_j -t\phi_j^{(+)}=E_j\phi_j^{(+)}. \label{Eq-phi-}  
\end{align}  
\end{subequations}  
These two equations are put together to provide single second-order differential equations for   
each of $\phi^{(\pm)}_j$;   
\begin{subequations}  
\label{Bessel(pm)}  
 \begin{align}  
   \frac{d^2\phi_j^{(-)}}{dr^2} + \frac{1}{r}\frac{d\phi_j^{(-)}}{dr}+  
      \Bigr(E_j^2-t^2-\frac{1}{r^2}(j-\frac12)^2\Bigr)\phi_j^{(-)} & =0, \quad {\rm if}\;\, E_j\neq t,   
    \label{Bessel-}\\  
   \frac{d^2\phi_j^{(+)}}{dr^2} + \frac{1}{r}\frac{d\phi_j^{(+)}}{dr}+  
      \Bigr(E_j^2-t^2-\frac{1}{r^2}(j+\frac12)^2\Bigr)\phi_j^{(+)} & =0, \quad {\rm if}\;\, E_j\neq -t.  
    \label{Bessel+}  
  \end{align}  
\end{subequations}  
  
If $E_j^2-t^2>0$,  these equations are Bessel differential equations,   
which are solved in terms of cylindrical functions.   
Putting   
\begin{equation}  
   \beta_j=\sqrt{E_j^2-t^2}, \quad |E_j|>|t|,   
\end{equation}  
we obtain the solutions to \eqref{Bessel-} and \eqref{Bessel+}  
\begin{equation}  
    \phi_j^{(-)}=C_1 J_{j-\frac12}(\beta_j r), \quad \phi_j^{(+)}=C_2 J_{j+\frac12}(\beta_j r),   
\end{equation}  
respectively, where  $C_1,C_2$ are constants.   
We here remark that Neumann functions have been deleted because of the boundary condition that   
$\phi_j^{(\pm)}$ should be bounded as $r\to 0$.   
We note further that the constants $C_1$ and $C_2$ are related to each other, since $\phi^{(\pm)}_j$   
are coupled through \eqref{Eq-phis}.    
From \eqref{Eq-phi-} together with the formula for Bessel functions    
\begin{equation}   
     J'_{\nu}(x)=\frac{\nu}{x}J_{\nu}(x)-J_{\nu+1}(x),   
\end{equation}  
we obtain the relation between $C_1$ and $C_2$   
\begin{equation}  
    (E_j+t)C_2=iC_1\beta_j.   
\end{equation}   
Since $\beta_j=\sqrt{E_j^2-t^2}>0$ with $|E_j|>|t|$, the above relation leads to   
\begin{subequations}  
 \begin{align}  
     \frac{C_1}{\sqrt{E_j+t}}=\frac{C_2}{i\sqrt{E_j-t}} \quad {\rm for} \quad E_j>0, \\  
      \frac{C_1}{\sqrt{|E_j+t|}}=\frac{C_2}{-i\sqrt{|E_j-t|}} \quad {\rm for} \quad E_j<0.   
 \end{align}  
\end{subequations}  
We thus find that solutions to $\hat{H}_t\Phi_j=E_j\Phi$ with $|E_j|>|t|$ take the form   
\begin{subequations}  
\label{sol |E|>|t|}  
 \begin{align}   
   \Phi_j(r,\theta)= & c\begin{pmatrix} \sqrt{E_j+t}e^{i(j-\frac12)\theta}J_{j-\frac12}(\beta_j r) \\  
                                               i\sqrt{E_j-t}e^{i(j+\frac12)\theta}J_{j+\frac12}(\beta_j r)   
                          \end{pmatrix} \quad {\rm for} \quad E_j>0,  \\  
   \Phi_j(r,\theta)= & c'\begin{pmatrix} \sqrt{|E_j+t|}e^{i(j-\frac12)\theta}J_{j-\frac12}(\beta_j r) \\  
                                               -i\sqrt{|E_j-t|}e^{i(j+\frac12)\theta}J_{j+\frac12}(\beta_j r)   
                          \end{pmatrix} \quad {\rm for} \quad E_j<0,  
 \end{align}  
\end{subequations}  
where $c$ and $c'$ are complex constants.    
   
If $E_j^2-t^2<0$,   
we may put Eq.~\eqref{Bessel-} and \eqref{Bessel+} in the form   
\begin{subequations}  
\label{mod Bessel eqs}  
 \begin{align}  
   \frac{d^2\phi_j^{(-)}}{dr^2} + \frac{1}{r}\frac{d\phi_j^{(-)}}{dr}-  
      \Bigr(t^2-E_j^2+\frac{1}{r^2}(j-\frac12)^2\Bigr)\phi_j^{(-)} & =0, \quad {\rm if}\;\, E_j\neq t,   
    \label{mod-Bessel-}\\  
   \frac{d^2\phi_j^{(+)}}{dr^2} + \frac{1}{r}\frac{d\phi_j^{(+)}}{dr}-  
      \Bigr(t^2-E_j^2 +\frac{1}{r^2}(j+\frac12)^2\Bigr)\phi_j^{(+)} & =0, \quad {\rm if}\;\, E_j\neq -t.  
    \label{mod-Bessel+}  
  \end{align}  
\end{subequations}  
These are the modified Bessel differential equations. Thus,    
if $|E_j|<|t|$, solutions take the form    
\begin{equation}  
        \phi_j^{(-)}(r)=C_1 I_{j-\frac12}(\varepsilon_j r), \quad \phi_j^{(+)}(r)=C_2 I_{j+\frac12}(\varepsilon_j r),   
\end{equation}  
where $I_{j\pm\frac12}$ denote the modified Bessel functions and where   
\begin{equation}  
    \varepsilon_j=\sqrt{t^2-E_j^2}, \qquad |E_j|<|t|.   
\end{equation}    
We note here that the modified Bessel function of the second kind have been deleted because of the   
boundary condition  that $\phi^{(\pm)}_j$ should be bounded as $r\to 0$.    
  
From \eqref{Eq-phi-} together with the formula for the modified Bessel functions   
\begin{equation}   
     I'_{\nu}(x)=\frac{\nu}{x}I_{\nu}(x)+I_{\nu+1}(x),   
\end{equation}  
we obtain the relation between $C_1$ and $C_2$   
\begin{equation}  
    (E_j+t)C_2=-iC_1\varepsilon_j.   
\end{equation}  
Since $\varepsilon_j=\sqrt{t^2-E_j^2}>0$ with $|t|>|E_j|$, this relation leads to   
\begin{subequations}  
 \begin{align}  
     \frac{C_1}{\sqrt{t+E_j}}=\frac{C_2}{-i\sqrt{t-E_j}} \quad {\rm for} \quad t>0, \\  
      \frac{C_1}{\sqrt{|t+E_j|}}=\frac{C_2}{i\sqrt{|t-E_j|}} \quad {\rm for} \quad t<0.   
 \end{align}  
\end{subequations}  
Thus, solutions to $\hat{H}_t\Phi_j=E_j\Phi_j$ with $|E_j|<|t|$ turns out to take the form   
\begin{subequations}  
\label{sol |E|<|t|}  
 \begin{align}   
   \Phi_j(r,\theta)= & c\begin{pmatrix} \sqrt{t+E_j}e^{i(j-\frac12)\theta}I_{j-\frac12}(\varepsilon_j r) \\  
                                               -i\sqrt{t-E_j}e^{i(j+\frac12)\theta}I_{j+\frac12}(\varepsilon_j r)   
                          \end{pmatrix} \quad {\rm for} \quad t>0,  \\  
   \Phi_j(r,\theta)= & c'\begin{pmatrix} \sqrt{|t+E_j|}e^{i(j-\frac12)\theta}I_{j-\frac12}(\varepsilon_j r) \\  
                                               i\sqrt{|t-E_j|}e^{i(j+\frac12)\theta}I_{j+\frac12}(\varepsilon_j r)   
                          \end{pmatrix} \quad {\rm for} \quad t<0,  
 \end{align}  
\end{subequations}  
where $c$ and $c'$ are complex constants.     
  
\begin{figure}[htbp]  
\begin{center}  
\includegraphics[width=0.4\columnwidth]{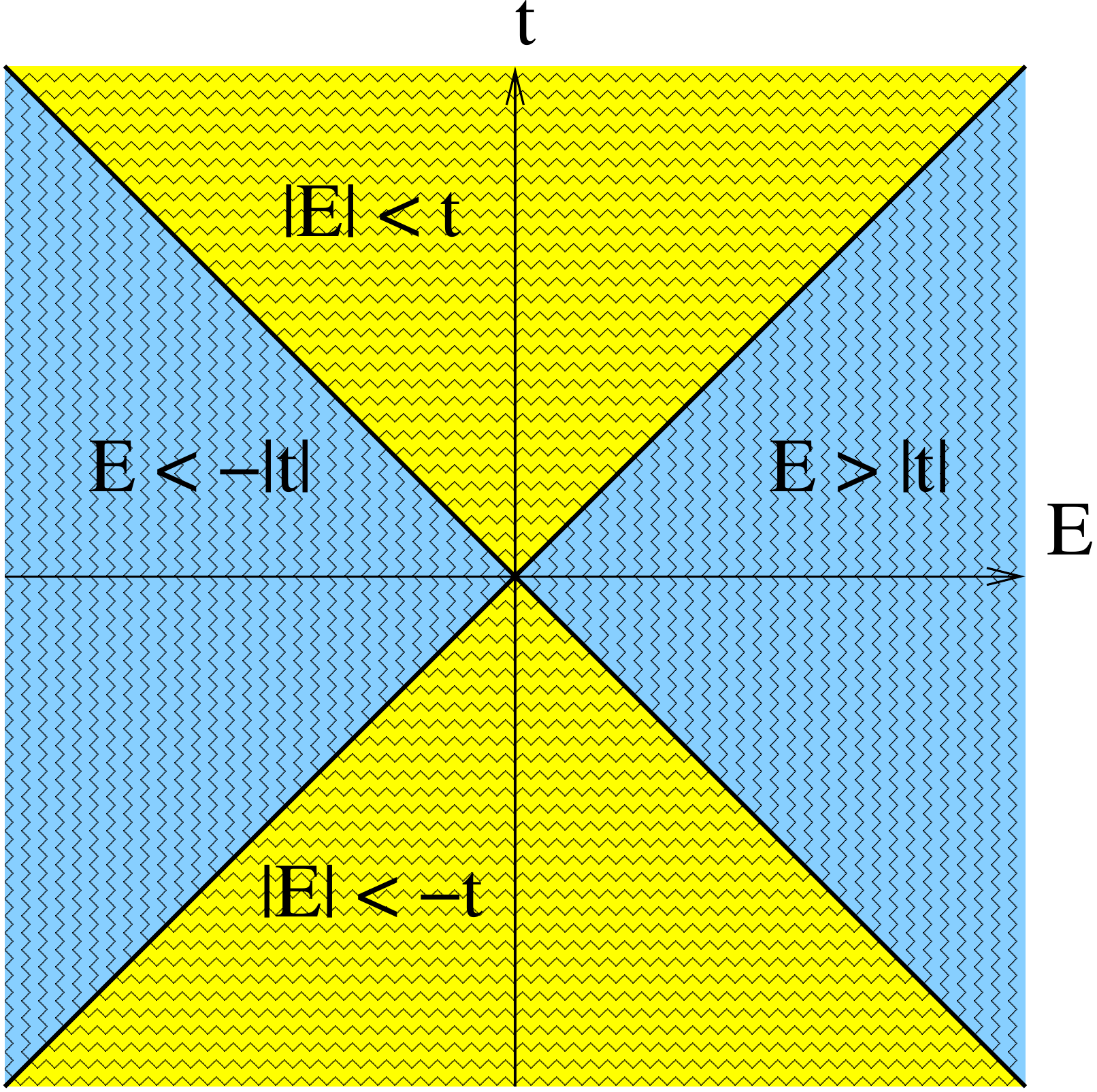}  
\end{center}  
\caption{Splitting of the $E$-$t$ plane into domains with $|E|>|t|$  
and with $|E|<|t|$. Each  domain consists of two connected components.    
To respective domains, feasible solutions \eqref{sol |E|>|t|} and \eqref{sol |E|<|t|}   
are assigned.   
\label{Et_plane} }  
\end{figure}  
  
\section{ Search for boundary conditions}  
\label{boundary conditions}  
\par  
To determine energy eigenvalues, we need a boundary condition for $\Phi_j(r,\theta)$.      
We assume that the initial eigenvalue equation $\hat{H}_t\Phi_j=E_j\Phi_j$ is defined on the disk   
$D^2_R$ of radius $R$, taking into account the $SO(2)$ symmetry of the Hamiltonian.    
If we are allowed to treat a single second-order differential equation,   
either \eqref{Bessel-} or \eqref{Bessel+} for $E_j^2-t^2>0$ and   
either \eqref{mod-Bessel-} or \eqref{mod-Bessel+} for $E_j^2-t^2<0$,    
we can pose the Robin boundary condition, for example,    
\begin{equation}  
\label{Robins}  
    \dot{\phi}^{(+)}_j(R)=\mu\phi^{(+)}_j(R) \quad{\rm or}\quad   
     \dot{\phi}^{(-)}_j(R)=\mu\phi^{(-)}_j(R), \quad \mu>0,   
\end{equation}  
where the superscript dot means the directional derivative with respect to the outward unit normal vector   
to the boundary $r=R$ of the disk.         
We note here that if $\mu=0$, the condition becomes a Neumann condition, and if $\mu\to +\infty$,   
the condition tends to a Dirichlet condition.   
However, if we simultaneously require both conditions given in \eqref{Robins}, we encounter a contradiction.     
We need a boundary condition for the coupled first-order equations, in place of the   
boundary condition for single second-order differential equations.   
  
\subsection{ Boundary conditions for Dirac equations}  
In the paper \cite{Asorey2013},   
boundary conditions are discussed to make self-adjoint the Dirac operator acting on spinors   
defined on a domain with boundary.   
  
To get an idea of boundary conditions, we start with the Green formula for the Laplacian    
\begin{equation}   
    \int_V \Bigl(u \nabla^2 v - v\nabla^2 u \Bigr)dV =   
    \int_S \Bigl(u\frac{\partial v}{\partial n}-v\frac{\partial u}{\partial n}\Bigr)dS,   
\end{equation}  
where $V$ is a domain bonded by the surface $S$, and where $\frac{\partial u}{\partial n}$ and   
$\frac{\partial v}{\partial n}$ denote the directional derivatives of $u$ and $v$ with respect to the outward    
unit normal $\boldsymbol{n}$ to the boundary $S$, respectively.   
Introducing the inner product for scalar functions on the domain $V$ and on the boundary by   
\begin{equation}   
     \langle u,v\rangle_V \quad {\rm and} \quad \langle \phi, \psi \rangle_S,   
\end{equation}   
respectively, we rewrite the Green formula as   
\begin{equation}   
    \langle u,\nabla^2 v \rangle_V -  \langle \nabla^2 u, v \rangle_V=   
    \langle u|_S, \dot{v}|_S \rangle_S - \langle \dot{u}|_S, v|_S\rangle_S,   
\end{equation}  
where $\dot{u}$ and $\dot{v}$ denote the directional derivatives of $u$ and $v$ with respect to the outward    
unit normal $\boldsymbol{n}$, respectively, and where  $\dot{u}|_S$ and $\dot{v}|_S$ denote   
$\dot{u}$ and $\dot{v}$ evaluated on $S$, respectively.   
From this equation we see that for $C^2$ functions satisfying the Robin boundary condition   
the right-hand side of the above equation vanishes, so that   
the Laplace operator $\nabla^2$ becomes symmetric.    
  
We wish to obtain a similar formula for our Dirac Hamiltonian $\hat{H}_t$.   
After \cite{Asorey2013}, we treat the Dirac operator on $\mathbb{R}^d$ which takes the form   
\begin{equation}  
\label{Dirac-Ham}   
   H=-i\sum_{j=1}^d \gamma^j\nabla_j + m \gamma^{d+1},   
\end{equation}  
where $\nabla_j=\partial/\partial x_j$ and where the gamma matrices satisfy   
\begin{subequations}  
 \begin{align}   
      \gamma^k\gamma^j+\gamma^j\gamma^k & =2 \delta^{jk}, \quad j,k=1,\dots,d, \\  
      \gamma^k\gamma^{d+1}+\gamma^{d+1}\gamma^k & =0, \\  
         (\gamma^{d+1})^2 & =\boldsymbol{1}, \\  
        (\gamma^{\nu})^{\dagger} & =\gamma^{\nu}, \quad \nu=1,\dots,d,d+1.   
 \end{align}  
\end{subequations}  
The inner products for multi-component functions on $V$ and on $S$ are defined as usual to be   
\begin{equation}  
    \langle \Phi, \Psi \rangle_V=\int_V \sum \overline{\Phi}_{\alpha}\Psi_{\alpha}dV, \quad   
    \langle \phi, \psi \rangle_S=\int_S \sum \overline{\phi}_{\alpha}\psi_{\alpha}dS,   
\end{equation}  
respectively.     
In order to obtain a formula which serves as a Green formula for the Dirac operator, we need   
the formula   
\begin{equation}  
\label{from-Gauss}   
     \int_V \vec{\nabla} f \,dV = \int_S f\vec{n} dS,   
\end{equation}  
where $f$ is a scalar function and $\vec{n}$ denotes the outward unit normal to $S$.   
If $d=3$, this formula is a consequence of the well-known Gauss divergence theorem.   
  
We now calculate the difference between $\langle \Phi,H\Psi\rangle_V$ and $\langle H\Phi,\Psi\rangle_V$.   
Since $\langle \Phi, \gamma^{d+1}\Psi\rangle = \langle \gamma^{d+1}\Phi, \Psi\rangle$,   
we have only to treat the differential operator terms.    
Adopting the summation convention and using \eqref{from-Gauss},  we can verify that    
\begin{equation}  
    \int_V \Bigl(\overline{\Phi}_{\alpha}(-i)\gamma^j_{\alpha \beta}\nabla_j\Psi_{\beta}   
                 -i \overline{\gamma}^j_{\alpha \beta}\nabla_j\overline{\Phi}_{\beta}\Psi_{\alpha} \Bigr)dV  
    =   
    -i\int_S \gamma^j_{\alpha \beta} n^j \overline{\Phi}_{\alpha}\Psi_{\beta}\, dS.   
\end{equation}  
Thus, the Green formula for the Dirac operator is  put in the form    
\begin{equation}  
\label{Dirac-Green}   
    \langle \Phi, H \Psi\rangle_V - \langle H \Phi, \Psi \rangle_V =   
    -i \langle \phi, \vec{\gamma}\cdot \vec{n} \,\psi \rangle_S,   
\end{equation}   
where $\phi=\Phi|_S, \psi=\Psi|_S$ and $\vec{\gamma}\cdot \vec{n}=\sum\gamma^j n^j$.   
  
Boundary conditions should be determined so that the boundary integral   
$\langle \phi, \vec{\gamma}\cdot \vec{n} \,\psi \rangle_S$ may vanish.   
According to \cite{Asorey2013}, the idea for a boundary condition is as follows:    
Let $K$ be any self-adjoint operator on the Hilbert space $\mathcal{H}(S)$ of multi-component functions  
on $S$, which is assumed to have the properties,   
\begin{subequations}  
 \begin{gather}  
                      K^2>0, \\  
          \vec{\gamma}\cdot \vec{n}\,K=-K\,\vec{\gamma}\cdot \vec{n}. \label{anti-commuting}  
 \end{gather}  
\end{subequations}   
Then, the $\mathcal{H}(S)$ splits into the orthogonal direct sum   
$\mathcal{H}(S)=\mathcal{H}^+(S)\oplus \mathcal{H}^-(S)$,  
where $\mathcal{H}^{(\pm)}(S)$ are subspaces such that   
$K|_{\mathcal{H}^+(S)}>0$ and $K|_{\mathcal{H}^-(S)}<0$.   
Since $K$ is self-adjoint and since positive eigenvalues and negative eigenvalues never coincide,   
$\mathcal{H}^+(S)$ and  $\mathcal{H}^-(S)$ are orthogonal.   
From \eqref{anti-commuting} together with the formula   
\begin{equation}  
\label{(gamma n)^2}   
          (\vec{\gamma}\cdot\vec{n})^2=\1,   
\end{equation}   
we can verify that   
\begin{equation}  
\label{+to-,-to+}   
    (\vec{\gamma}\cdot\vec{n})\mathcal{H}^{(\pm)}(S)=\mathcal{H}^{(\mp)}(S).   
\end{equation}   
From \eqref{Dirac-Green} and \eqref{+to-,-to+}, it then turns out that if   
\begin{equation}   
\label{tentative bdy cond}  
    \Phi|_S,\Psi|_S \in \mathcal{H}^{(-)}(S), \;\mbox{or} \;    
   \Phi|_S,\Psi|_S \in \mathcal{H}^{(+)}(S),   
\end{equation}   
then   
\begin{equation}  
   \langle \Phi,H\Psi\rangle_S-\langle H\Phi, \Psi \rangle_S =0,   
\end{equation}   
With some Sobolev conditions on $\Psi$'s, the $H$ becomes self-adjoint.     
However, we wish to stress that the condition \eqref{anti-commuting} is an sufficient condition for   
\eqref{+to-,-to+} and can be relaxed, as will be done in the following subsection.   
  
\subsection{An APS boundary condition on the sphere}  
\par  
We specialize the domain $S$ to the sphere $S=S^{d-1}$ of radius $R$, but relax  
the condition \eqref{anti-commuting}.      
Let $\vec{n}$ be the outward unit normal vector field on $S^{d-1}$ and   
$\vec{t}_a$ with $a=1,\dots,d-1$ be (locally-defined) orthonormal tangent vector fields on $S^{d-1}$.   
Then, one has   
\begin{equation}   
\label{decomp of unity}  
       |\vec{n}\rangle \langle \vec{n}| + \sum_a |\vec{t}_a\rangle \langle \vec{t}_a| = \1.   
\end{equation}   
Though $\vec{t}_a$ are defined locally, the above equation can be extended globally on the sphere.   
This is because on the intersection of the domains of $\vec{t}_a$ and of $\vec{t}_a{}'$ they are related   
by  $\vec{t}_a{}'=\sum g_{ba}\vec{t}_b$ with $(g_{ba})\in SO(n-1)$ and hence the equality   
$\sum |\vec{t}_a\rangle \langle \vec{t}_a|=\sum |\vec{t}_a{}'\rangle \langle \vec{t}_a{}'|$ holds.   
By using \eqref{decomp of unity}, we decompose the operator   
$\langle \vec{\gamma}|\vec{\nabla}\rangle=\sum \gamma^j\nabla_j$   
into the sum of radial and tangential components;   
\begin{equation}   
 \langle\vec{\gamma}| \vec{\nabla}\rangle  =     
   \langle \vec{\gamma}|\vec{n}\rangle \langle\vec{n}|\vec{\nabla}\rangle +   
        \sum_a  \langle \vec{\gamma}|\vec{t}_a\rangle \langle \vec{t}_a|\vec{\nabla}\rangle .  
        \label{rad-tan-decomp}  
\end{equation}  
Since $\vec{n}=\frac{1}{r}\vec{r}$, we have   
\begin{equation}   
\label{rad-comp}  
    \langle\vec{n}|\vec{\nabla}\rangle=\sum_j \frac{x_j}{r}\frac{\partial}{\partial x_j} =   
    \frac{\partial}{\partial r}.   
\end{equation}  
We here introduce the tangential component $\gamma_a$ of $\vec{\gamma}$ and   
the tangential operator (tangent vector to $S^{d-1}$) $X_a$ by   
\begin{equation}  
\label{tan-comp}   
     \gamma_a = \langle \vec{\gamma} | \vec{t}_a \rangle, \quad   
     \frac{1}{r} X_a =\langle \vec{t}_a| \vec{\nabla} \rangle, \quad a=1,\dots,d-1,   
\end{equation}   
respectively.   
From \eqref{rad-tan-decomp}, \eqref{rad-comp}, and \eqref{tan-comp},   
we obtain the decomposition    
\begin{equation}   
    \langle\vec{\gamma}| \vec{\nabla}\rangle = \langle \vec{\gamma}|\vec{n}\rangle \partial_r +   
       \frac{1}{r}\sum_a \gamma_a X_a .   
\end{equation}  
  
We now show that the tangential operator $\sum \gamma_a X_a$ is not anti-commuting with    
$\langle\vec{\gamma}|\vec{n}\rangle$.   
To this end, we start with the formulas,   
\begin{subequations}   
\begin{align}   
  \langle\vec{\gamma}|\vec{n}\rangle \gamma_a + \gamma_a \langle\vec{\gamma}|\vec{n}\rangle=0, \\  
   X_a|\vec{n}_\rangle =|\vec{t}_a\rangle,   
\end{align}  
\end{subequations}   
which are verified, respectively, as follows:   
\begin{subequations}  
\begin{align}      
   &   \langle\vec{\gamma}|\vec{n}\rangle \gamma_a + \gamma_a \langle\vec{\gamma}|\vec{n}\rangle    
  =  \gamma^j n^j t^k_a\gamma^k + t^k_a \gamma^k \gamma^jn^j   
  =  2\delta^{jk}\boldsymbol{1}n^jt^k_a \,=\, 0, \\  
  & \frac{1}{r}X_a|\vec{n}\rangle =\sum t_a^j\frac{\partial}{\partial x_j}\bigl(\frac{x_k}{r}\bigr)=   
  \frac{t_a^j}{r^2}\bigl(\delta_{kj}r -\frac{x_kx_j}{r}\bigr)=\frac{t_a^k}{r}.  
\end{align}  
\end{subequations}  
Hence, we verify that   
\begin{align}   
   \langle\vec{\gamma}|\vec{n}\rangle \sum_a \gamma_a X_a +   
      \sum_a \gamma_a X_a \langle\vec{\gamma}|\vec{n}\rangle & =   
   \sum_a\bigl(\langle\vec{\gamma}|\vec{n}\rangle \gamma_a +   
              \gamma_a \langle\vec{\gamma}|\vec{n}\rangle\bigr)X_a   
   +\sum \gamma_a \langle \vec{\gamma}|\vec{t}_a\rangle  \nonumber \\  
  & = \sum_a \gamma_a^2=(d-1)\1.  \label{gamma n gamma X}  
\end{align}  
  
Now, we put the Hamiltonian \eqref{Dirac-Ham} in the form   
\begin{equation}   
   H=-i \langle \vec{\gamma}|\vec{n}\rangle \partial_r -\frac{i}{r}\sum_a \gamma_a X_a + m\gamma^{d+1}.   
\end{equation}   
Picking up the second and the third terms from the right-hand side of the above equation   
and restricting  them to $S^{d-1}$ of radius $R$,   
we define the operator     
\begin{equation}  
\label{def-A(m)}   
   A(\mu)=-\frac{i}{R}\sum \gamma_a X_a +\mu \gamma^{d+1},   
\end{equation}   
where the mass parameter $m$ has been replaced by the parameter $\mu$   
which may take negative values.     
Through a straightforward procedure together with the formula  
\begin{equation}   
  \gamma_a\gamma^{d+1}+\gamma^{d+1}\gamma_a  =0,    
\end{equation}    
we can verify that   
\begin{subequations}  
\label{A(m)}  
\begin{gather}   
      A(\mu)^2\geq \mu^2\1 >0, \\  
  \langle \vec{\gamma}|\vec{n}\rangle A(\mu) = -A(\mu) \langle \vec{\gamma}|\vec{n}\rangle   
    -\frac{i}{R}(d-1)\1,  
\end{gather}    
\end{subequations}  
where the last equation is a consequence of \eqref{gamma n gamma X}.   
  
We may take $K$ as   
\begin{equation}  
\label{K(mu)}   
     K(\mu)=i \langle \vec{\gamma}|\vec{n}\rangle A(\mu), \quad \mu\neq 0.   
\end{equation}   
Then, from \eqref{A(m)} together with $\langle \vec{\gamma}|\vec{n}\rangle^2=\boldsymbol{1}$,   
we can show that   
\begin{subequations}  
\begin{gather}   
      K(\mu)^2=A(\mu)^2 +\frac{d-1}{R}K(\mu), \label{K(mu)^2}\\  
  \langle \vec{\gamma}|\vec{n}\rangle K(\mu) =   
   -K(\mu) \langle \vec{\gamma}|\vec{n}\rangle +\frac{d-1}{R}\langle \vec{\gamma}|\vec{n}\rangle.   
   \label{K(mu)gamma(n)}  
\end{gather}    
\end{subequations}  
Eq.~\eqref{K(mu)^2} implies that $K(\mu)$ has no zero eigenvalue for $\mu\neq 0$.   
Because of \eqref{K(mu)gamma(n)}, the theory of Asorey {\it et al} \cite{Asorey2013}   
does not apply in its original form (see \eqref{anti-commuting}).   
  
Though the operator $K(\mu)$ does not anti-commute with $\langle \vec{\gamma}|\vec{n}\rangle$,   
we may use the $K(\mu)$ in order to describe a boundary condition.   
This is because the operator $\gamma_n^{}=\langle \vec{\gamma}|\vec{n}\rangle$ maps   
$\mathcal{H}^{(+)}(S^{d-1})$ to $\mathcal{H}^{(-)}(S^{d-1})$ and vice versa, as is shown below,   
where  $\mathcal{H}^{(+)}(S^{d-1})$ and $\mathcal{H}^{(-)}(S^{d-1})$ are defined to be subspaces on which   
one has $K(\mu)>0$ and $K(\mu)<0$, respectively.       
Let $\phi^{(+)}$ be an eigenstate associated with a positive eigenvalue $\kappa$ of $K(\mu)$.   
Then one has $K(\mu)\phi^{(+)}=\kappa \phi^{(+)},\,\kappa>0$.   
Operating the both side of this equation with $\gamma_n^{}$ and using \eqref{K(mu)gamma(n)},   
we obtain   
\begin{equation}   
\label{K(mu) gamma n}  
    K(\mu)\gamma_n^{}\phi^{(+)} = -\bigl(\kappa -\frac{d-1}{R}\bigr)\gamma_n^{}\phi^{(+)}.   
\end{equation}   
This implies that if $\kappa-\frac{d-1}{R}>0$ then $\gamma_n^{}\phi^{(+)}$ has an eigenstate   
associated with a negative eigenvalue.   
It then follows that if the smallest positive eigenvalue satisfies $\kappa-\frac{d-1}{R}>0$ then   
the operator $\gamma_n^{}=\langle \vec{\gamma}|\vec{n}\rangle$ maps   
$\mathcal{H}^{(+)}(S^{d-1})$ to $\mathcal{H}^{(-)}(S^{d-1})$.   
In a similar manner, we can show that operator $\gamma_n^{}$ maps   
$\mathcal{H}^{(-)}(S^{d-1})$ to $\mathcal{H}^{(+)}(S^{d-1})$.    
This property is sufficient for the Dirac operator to be self-adjoint. In fact,  from   
\eqref{Dirac-Green} and the above boundary condition,  we obtain   
\begin{equation}  
   \langle \Phi,H\Psi\rangle_{B^d}-\langle H\Phi,\Psi\rangle_{B^d}=0.   
\end{equation}   
where $B^d$ is the ball whose boundary is the sphere $S^{d-1}$ of radius $R$.   
The boundary condition is now described as     
\begin{equation}  
         \Psi|_{S^{d-1}}\in \mathcal{H}^{(-)}(S^{d-1}),\quad {\rm or} \quad   
         \Psi|_{S^{d-1}}\in \mathcal{H}^{(+)}(S^{d-1}),   
\end{equation}   
which we call the APS boundary condition after \cite{APS}.    
   
\subsection{ The APS boundary condition for $\hat{H}_t$}  
\par   
We now return to our initial eigenvalue problem with the Hamiltonian $\hat{H}_t$ on the   
disk $D^2_R$ of radius $R$.   
From \eqref{polar-Ham}, we see that the operator $A(t)$ corresponding to \eqref{def-A(m)}   
is expressed as   
\begin{equation}   
   A(t)=-\frac{i}{R}\sigma_{\theta}\partial_{\theta} + t\sigma_3,    
\end{equation}  
and then the operator $K_t$ is defined from \eqref{K(mu)} to be   
\begin{align}   
     K_t= i\sigma_r A(t)   
     =  \begin{pmatrix} \frac{i}{R}\frac{\partial}{\partial \theta} & -it e^{-i\theta} \\  
                                it e^{i\theta} & -\frac{i}{R}\frac{\partial}{\partial \theta} \end{pmatrix},   
\end{align}  
where $t\neq 0$ in the present section.  We will treat the case  of $t=0$ later.   
  
For the sake of confirmation, we show that $\sigma_r$ and $K_t$ do not anti-commute.   
Along with the fact that $\partial_{\theta} \sigma_r=\sigma_{\theta}$, a calculation provides   
$\sigma_rK_t+K_t\sigma_r = \frac{1}{R}\sigma_r$.   
  
To describe the APS boundary condition, we have to find the eigenvalues and the eigenstates for   
the operator $K_t$.   
As is seen form \eqref{eigen-st(j)},  the eigenstates associated with the eigenvalue $j$ are   
to be expressed as   
\begin{equation}   
     \phi_j(\theta)=   
    \begin{pmatrix} b_j e^{i(j-\frac12)\theta} \\  a_j e^{i(j+\frac12)\theta} \end{pmatrix}.   
\end{equation}  
We note here that the operator $K_t$ is rewritten as    
\begin{equation}   
     K_t=-\frac{1}{R}\sigma_3 J +\frac{1}{2R}\1+ t\sigma_{\theta}.   
\end{equation}  
With this in mind, we solve the eigenvalue problem for the operator $K_t-\frac{1}{2R}\1$;   
\begin{equation}   
     \Bigl(K_t-\frac{1}{2R}\1 \Bigr)\phi_j =\lambda_j \phi_j.    
\end{equation}   
Taking $J\phi_j=j\phi_j$ into account, we find that the above equation is reduced to   
the algebraic eigenvalue equation for $\lambda_j$,   
\begin{equation}  
\label{alg-eigenvalue eq}   
       \begin{pmatrix} -\frac{j}{R} & -it \\ it & \frac{j}{R} \end{pmatrix} \begin{pmatrix} b_j \\ a_j \end{pmatrix}   
         = \lambda_j \begin{pmatrix} b_j \\ a_j \end{pmatrix} .   
\end{equation}   
A straightforward calculation provides us with   
the eigenvalues of $K_t-\frac{1}{2R}\1$,      
\begin{equation}  
\label{lambda(pm)_j}   
    \lambda^{\pm}_j =\pm \sqrt{\frac{j^2}{R^2}+t^2}   
\end{equation}   
together with the associated eigenvectors, respectively,    
\begin{equation}  
\label{+- eigenvectors}  
    \begin{pmatrix} b_j \\ a_j \end{pmatrix}^{\!+} =  
            c_j \begin{pmatrix}  -it \\ \frac{j}{R}+\lambda_j^{+}  \end{pmatrix},  \quad  
    \begin{pmatrix} b_j \\ a_j \end{pmatrix}^{\!-} =  
            c'_j \begin{pmatrix} -it \\ \frac{j}{R}+\lambda_j^{-}  \end{pmatrix}, \quad t\neq 0.   
\end{equation}  
The eigenvalues of $K_t$ is then given by $\kappa^{\pm}_j:=\lambda^{\pm}_j+\frac{1}{2R}$.  Since   
\begin{equation}   
    \frac{1}{2R}+\lambda_j^- =\frac{1}{2R}-\sqrt{\frac{j^2}{R^2}+t^2}\leq   
     \frac{1}{2R}-\sqrt{\frac{1}{4R^2}+t^2} <0,  \quad j\in \bigl\{\pm\frac12, \pm\frac32,\cdots\bigr\},   
     \; t\neq 0,   
\end{equation}  
and since $\lambda_j^+ +\frac{1}{2R}>0$, we have obtained negative and positive eigenvalues $\kappa^{\pm}_j$   
of $K_t$, $t\neq 0$, together with the associated eigenstates,     
\begin{subequations}  
\label{bdy-sol}  
\begin{align}   
    \phi^{(-)}_j(\theta) = &  
   c'_j  \begin{pmatrix} -it\, e^{i(j-\frac12)\theta} \\   
                               (\frac{j}{R}+\lambda_j^{-}) e^{i(j+\frac12)\theta} \end{pmatrix}   
     \quad {\rm for} \quad \kappa^-_j:=\frac{1}{2R}+\lambda_j^- <0, \\                                  
     \phi^{(+)}_j(\theta) = &   
   c_j  \begin{pmatrix}   -it \,e^{i(j-\frac12)\theta} \\    
                                (\frac{j}{R}+\lambda_j^{+})e^{i(j+\frac12)\theta} \end{pmatrix} \quad {\rm for} \quad   
           \kappa^+_j:=\frac{1}{2R}+\lambda_j^+>0.  
\end{align}  
\end{subequations}  
Since $\kappa_{j}^{+}=\kappa^{+}_{-j}$ and $\kappa_{j}^{-}=\kappa^{-}_{-j}$, each eigenvalue is   
doubly degenerate. Put another way, the eigenstates $\phi^{(+)}_{\pm j}$ and  $\phi^{(-)}_{\pm j}$  
belong to the same eigenspaces, respectively.   
  
The spaces $\mathcal{H}^{(\pm)}(\partial D^2_R)$ are spanned by eigenstates   
$\phi^{(\pm)}_j, \,j\in\{\pm \frac12, \pm\frac23, \cdots\}$, respectively,  and then   
the total space $\mathcal{H}(\partial D^2_R)$ attached to the boundary $\partial D^2_R$   
is decomposed into the direct sum of these two subspaces;   
$\mathcal{H}(\partial D^2_R) = \mathcal{H}^{(+)}(\partial D^2_R) \oplus \mathcal{H}^{(-)}(\partial D^2_R)$.   
The orthogonality of $\mathcal{H}^{(+)}(\partial D^2_R)$ and $\mathcal{H}^{(-)}(\partial D^2_R)$   
is easy to prove, which reduces to the orthogonality of the two eigenvectors given in   
\eqref{+- eigenvectors}.  In addition,   
we can show that $\sigma_r\mathcal{H}^{(\mp)}(\partial D^2_R)=\mathcal{H}^{(\pm)}(\partial D^2_R)$   
by using Eq.~\eqref{K(mu) gamma n} with $d=2$.  In fact,  
from $K_t\phi^{(+)}_j=\kappa^+_j\phi^{(+)}_j$, we obtain   
$K_t\sigma_r\phi^{(+)}_j=-(\kappa^+_j -\frac{1}{R})\sigma_r\phi^{(+)}_j$.   
Since $\kappa^+_j-\frac{1}{R}=-\frac{1}{2R}+\sqrt{\frac{j^2}{R^2}+t^2}\geq -\frac{1}{2R}+\sqrt{\frac{1}{4R^2}+t^2}>0$   
for all $j$ and for $t\neq 0$,  the $\sigma_r\phi^{(+)}_j$ is an eigenstate associated with negative eigenvalue   
$-(\kappa^+_j-\frac{1}{R})$, which implies that   
$\sigma_r\mathcal{H}^{(+)}(\partial D^2_R) \subset \mathcal{H}^{(-)}(\partial D^2_R)$.    
Taking into account the fact $\sigma_r^2=\boldsymbol{1}$ further,  we  verify that   
$\sigma_r\mathcal{H}^{(-)}(\partial D^2_R)=\mathcal{H}^{(+)}(\partial D^2_R)$.   
In a similar manner, we verify that   
$\sigma_r\mathcal{H}^{(+)}(\partial D^2_R)=\mathcal{H}^{(-)}(\partial D^2_R)$.   
  
The APS boundary condition is thus expressed as   
\begin{equation}  
\label{APS Disk}  
    \Phi_j(R,\theta)\in \mathcal{H}^{(-)}(\partial D^2_R) \quad {\rm or} \quad   
    \Phi_j(R,\theta)\in \mathcal{H}^{(+)}(\partial D^2_R).   
\end{equation}

\section{ Edge states} \label{edge states}  
\par  
We are interested in eigenvalues of the full quantum Hamiltonian $\hat{H}_t$  under the APS boundary   
condition.   
According to whether the parameter pair $(E,t)$ belongs to the domain defined by $|E|<|t|$ or   
to that defined by $|E|>|t|$, feasible solutions take the form \eqref{sol |E|<|t|} or \eqref{sol |E|>|t|}.     
The APS boundary condition \eqref{APS Disk} is applied to those solutions separately.   
In this section, we treat the case of $|E_j|<|t|$ only, and postpone the case $|E_j|>|t|$ to a later section.    
  
We begin with the boundary condition $\Phi_j(R,\theta)\in \mathcal{H}^{(-)}(\partial D^2_R)$   
and will deal with the boundary condition $\Phi_j(R,\theta)\in \mathcal{H}^{(+)}(\partial D^2_R)$   
in the latter part of this section.   
According to the classification given in \eqref{sol |E|<|t|}, we treat two cases of $t>0$ and   
$t<0$ separately.     
From \eqref{sol |E|<|t|} and \eqref{bdy-sol},  the boundary condition   
$\Phi_j(R,\theta)\in \mathcal{H}^{(-)}(\partial D^2_R)$ for $t>0$ takes the form   
\begin{equation}   
\label{bdy-cond (-)}  
     \begin{pmatrix}   
     \sqrt{t+E_j}e^{i(j-\frac12)\theta}I_{j-\frac12}(\varepsilon_j R) \\  
                                               -i\sqrt{t-E_j}e^{i(j+\frac12)\theta}I_{j+\frac12}(\varepsilon_j R)   
     \end{pmatrix} = c \begin{pmatrix} -it\, e^{i(j-\frac12)\theta} \\   
                               (\frac{j}{R}+\lambda_j^{-}) e^{i(j+\frac12)\theta} \end{pmatrix} ,  
\end{equation}  
From the above equation, one obtains the equation for $E_j$ with $|E_j|<|t|$ and $t>0$,   
\begin{equation}   
    \frac{\sqrt{t+E_j}I_{j-\frac12}(\varepsilon_j R)}{-it}=  
    \frac{-i\sqrt{t-E_j}I_{j+\frac12}(\varepsilon_j R)}{\frac{j}{R}+\lambda^-_j}, \quad   
       \varepsilon_j=\sqrt{t^2-E_j^2},    
\end{equation}   
which is arranged as   
\begin{equation}  
\label{eigen-energy eq}   
    t\sqrt{\frac{t+E_j}{t-E_j}} I_{j-\frac12}(\varepsilon_j R) =   
    \Bigl(\frac{j}{R}+\sqrt{\frac{j^2}{R^2}+t^2}\Bigr) I_{j+\frac12}(\varepsilon_j R).   
\end{equation}   
If there is a solution to this equation, there exists an edge state satisfying the APS boundary condition.   
This is the case if $j>0$ and $E_j<0$ or if $j<0$ and $E_j>0$.   
In fact, if $j>0$ and $E_j<0$, one has   
\begin{equation}   
      0<t<\frac{j}{R}+\sqrt{\frac{j^2}{R^2}+t^2},   \quad  0< t\sqrt{\frac{t+E_j}{t-E_j}} <t,  
\end{equation}  
and further one has $I_{j-\frac12}(\varepsilon_j R)> I_{j+\frac12}(\varepsilon_j R)$ for $j>0$ \cite{IN},  
and hence Eq.~\eqref{eigen-energy eq} may have a solution.   
If $j<0$ and $E_j>0$, one has   
\begin{equation}   
    0< \frac{j}{R}+\sqrt{\frac{j^2}{R^2}+t^2}<t,  \quad 0<t<  t\sqrt{\frac{t+E_j}{t-E_j}},  
\end{equation}   
and $I_{j-\frac12}(\varepsilon_j R)< I_{j+\frac12}(\varepsilon_j R)$ for $j<0$ because of   
$I_{j-\frac12}=I_{-|j|-\frac12}=I_{|j|+\frac12}$ and $I_{j+\frac12}=I_{-|j|+\frac12}=I_{|j|-\frac12}$,   
so that Eq.~\eqref{eigen-energy eq} may have a solution, too.   
  
On the contrary, if $j>0$ and $E_j>0$, one has   
\begin{equation}   
    0<t<\frac{j}{R}+\sqrt{\frac{j^2}{R^2}+t^2},  \quad 0<t< t\sqrt{\frac{t+E_j}{t-E_j}},  
\end{equation}   
and further $I_{j-\frac12}(\varepsilon_j R)> I_{j+\frac12}(\varepsilon_j R)$.   
If  $j<0$ and $E_j<0$, one has   
\begin{equation}   
       0< \frac{j}{R}+\sqrt{\frac{j^2}{R^2}+t^2}<t,  \quad  0< t\sqrt{\frac{t+E_j}{t-E_j}} <t,   
\end{equation}  
and  $I_{j-\frac12}(\varepsilon_j R)< I_{j+\frac12}(\varepsilon_j R)$.   
In these two cases, Eq.~\eqref{eigen-energy eq} may have no solution for $E_j$.    
A graphical illustration of the existence or non-existence of solutions are shown in   
Fig.~\ref{Fig j>0,t>0, BesselI} for $j=11/2, R=1$ and $t=0.01,0.1,1$.   
The left-hand and right-hand sides of \eqref{eigen-energy eq} are described as (red and blue)   
functions of $E$, respectively.  The intersection of the (red and blue) graphs gives an eigen-energy,   
when projected onto the $E$-axis.      
  
\begin{figure}  
\begin{center}  
\includegraphics[width=0.3\columnwidth]{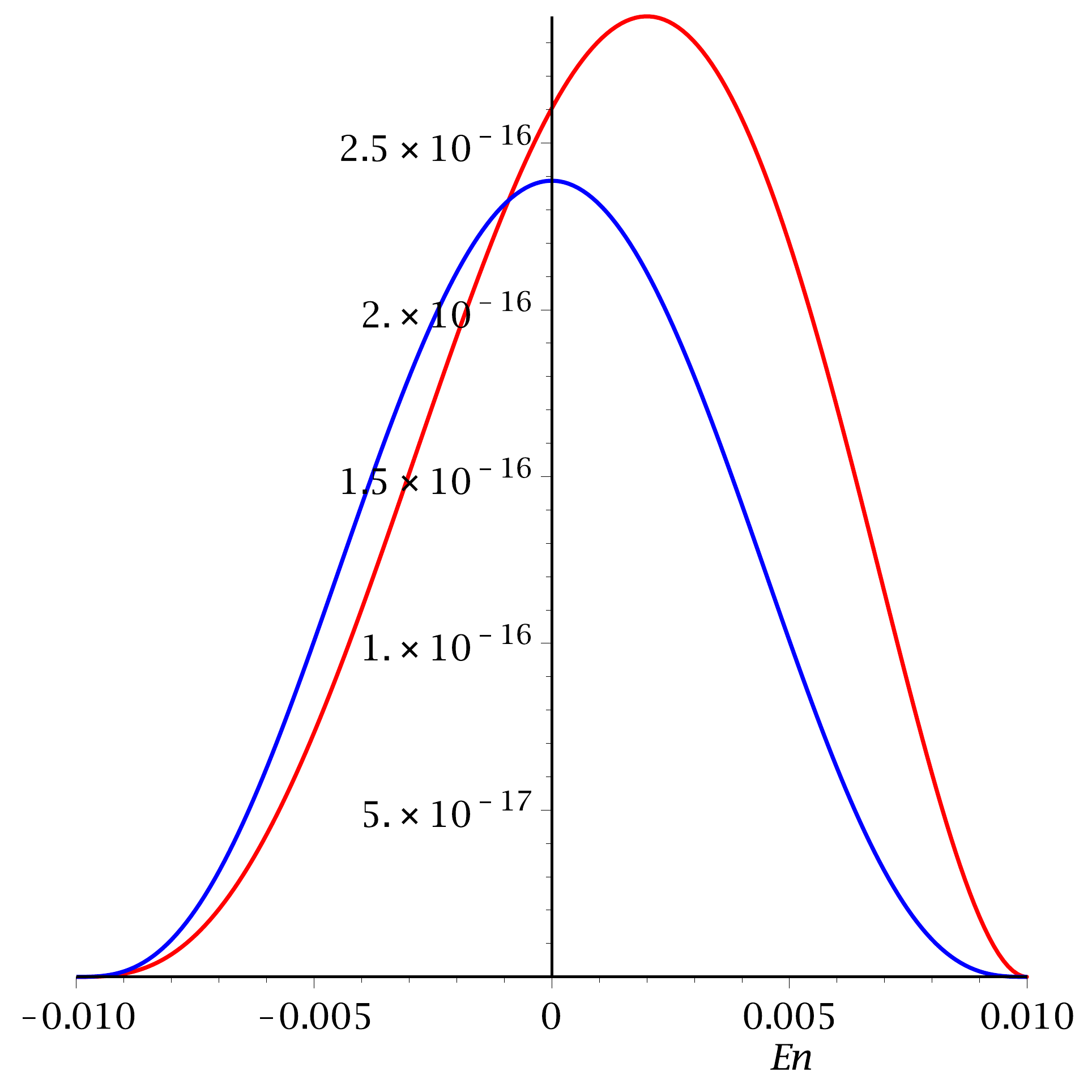}\quad  
\includegraphics[width=0.3\columnwidth]{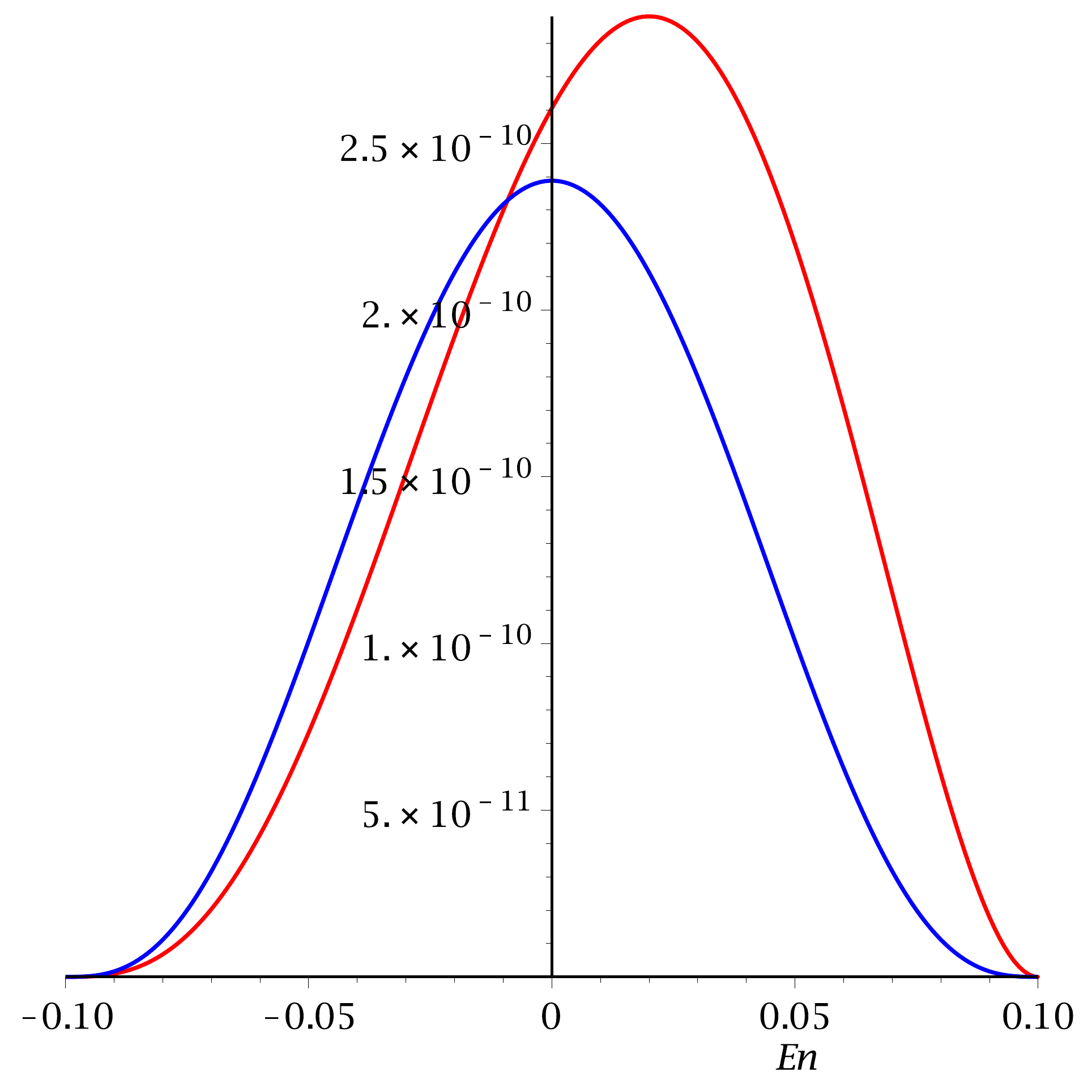} \quad  
\includegraphics[width=0.3\columnwidth]{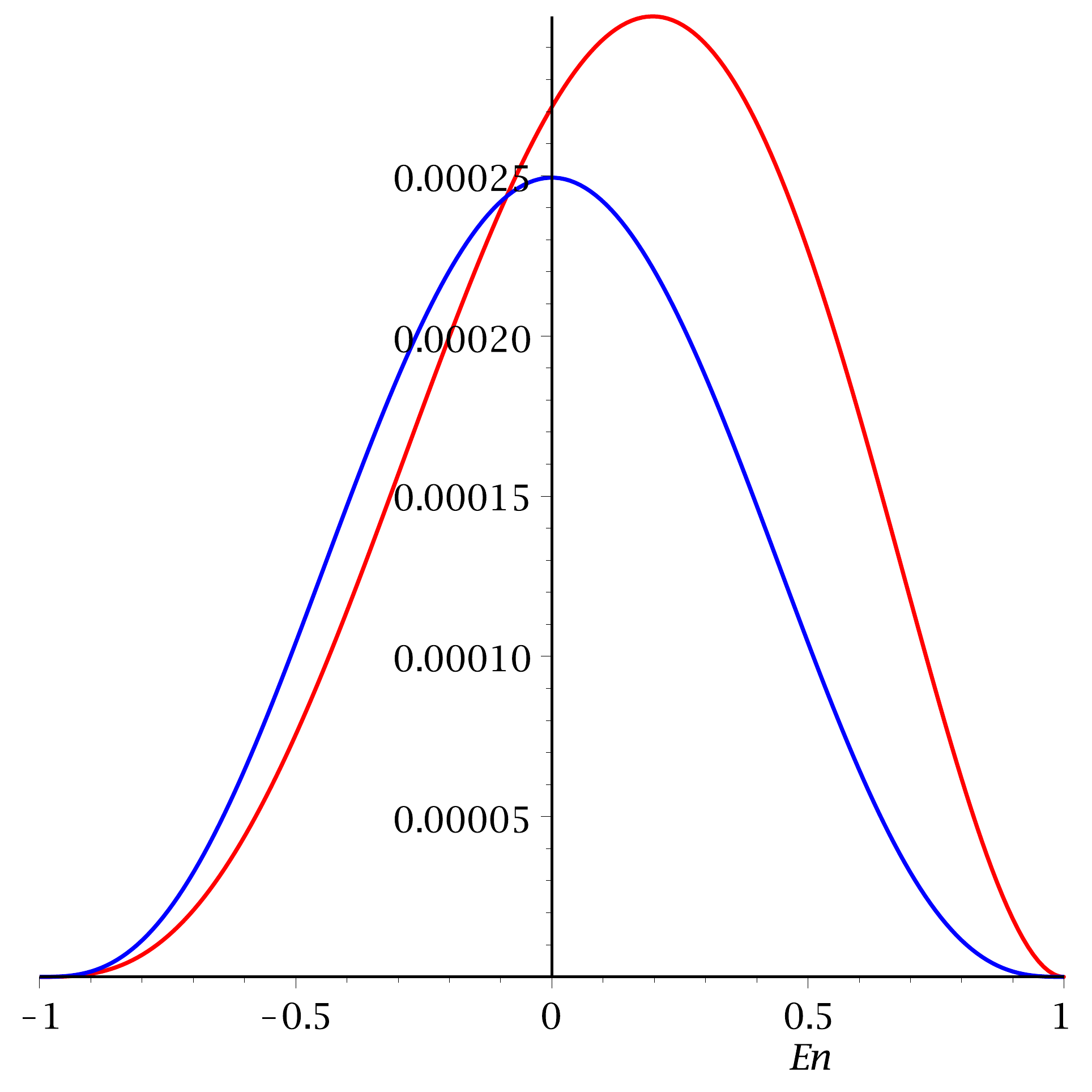} \quad  
\end{center}  
\caption{Graphical solution of equition (\protect{\ref{eigen-energy eq}}) for $j=11/2$, $R=1$  
and for $t=0.01, 0.1, 1$.}  
\label{Fig j>0,t>0, BesselI}  
\end{figure}  
  
We proceed to the case of $t<0$.   
For $t<0$, the boundary condition $\Phi_j(R,\theta)\in \mathcal {H}^{(-)}(\partial D^2_R)$   
takes the  form   
\begin{equation}   
     \begin{pmatrix}   
     \sqrt{|t+E_j|}e^{i(j-\frac12)\theta}I_{j-\frac12}(\varepsilon_j R) \\  
                                               i\sqrt{|t-E_j|}e^{i(j+\frac12)\theta}I_{j+\frac12}(\varepsilon_j R)   
     \end{pmatrix} = c \begin{pmatrix} -it\, e^{i(j-\frac12)\theta} \\   
                               (\frac{j}{R}+\lambda_j^{-}) e^{i(j+\frac12)\theta} \end{pmatrix} ,  
\end{equation}  
which is arranged as   
\begin{equation}  
\label{eigen-energy eq(t<0)}   
    |t|\sqrt{\frac{|t+E_j|}{|t-E_j|}} I_{j-\frac12}(\varepsilon_j R) =   
    \Bigl(\frac{j}{R}+\sqrt{\frac{j^2}{R^2}+t^2}\Bigr) I_{j+\frac12}(\varepsilon_j R).   
\end{equation}   
In contrast with the case of $t>0$, for $t<0$ there is a solution to this equation if $j>0$ and $E_j>0$   
or if $j<0$ and $E_j<0$.   
In fact, if  $j>0$ and $E_j>0$, one has   
\begin{equation}   
  0<|t|< \frac{j}{R}+\sqrt{\frac{j^2}{R^2}+t^2},  \quad  0<|t|\sqrt{\frac{|t+E_j|}{|t-E_j|}}<|t|,   
\end{equation}    
and further $I_{j-\frac12}(\varepsilon_j R)> I_{j+\frac12}(\varepsilon_j R)$,   
and if $j<0$ and $E_j<0$, then   
\begin{equation}   
 0< \frac{j}{R}+\sqrt{\frac{j^2}{R^2}+t^2}<|t|,   \quad 0<|t|<|t|\sqrt{\frac{|t+E_j|}{|t-E_j|}},   
\end{equation}    
and further $I_{j-\frac12}(\varepsilon_j R)< I_{j+\frac12}(\varepsilon_j R)$.   
  
On the contrary, if $j>0$ and $E_j<0$ of if $j<0$ and $E_j>0$, Eq.~\eqref{eigen-energy eq(t<0)}   
may have no solution by the same reasoning as that in the case of $t>0$.   
A graphical illustration is shown in Fig.~\ref{Fig j>0,t<0, BesselI} for $j=11/2, R=1$ and $t=-1,-0.1,-0.01$.   
  
\begin{figure}  
\begin{center}  
\includegraphics[width=0.3\columnwidth]{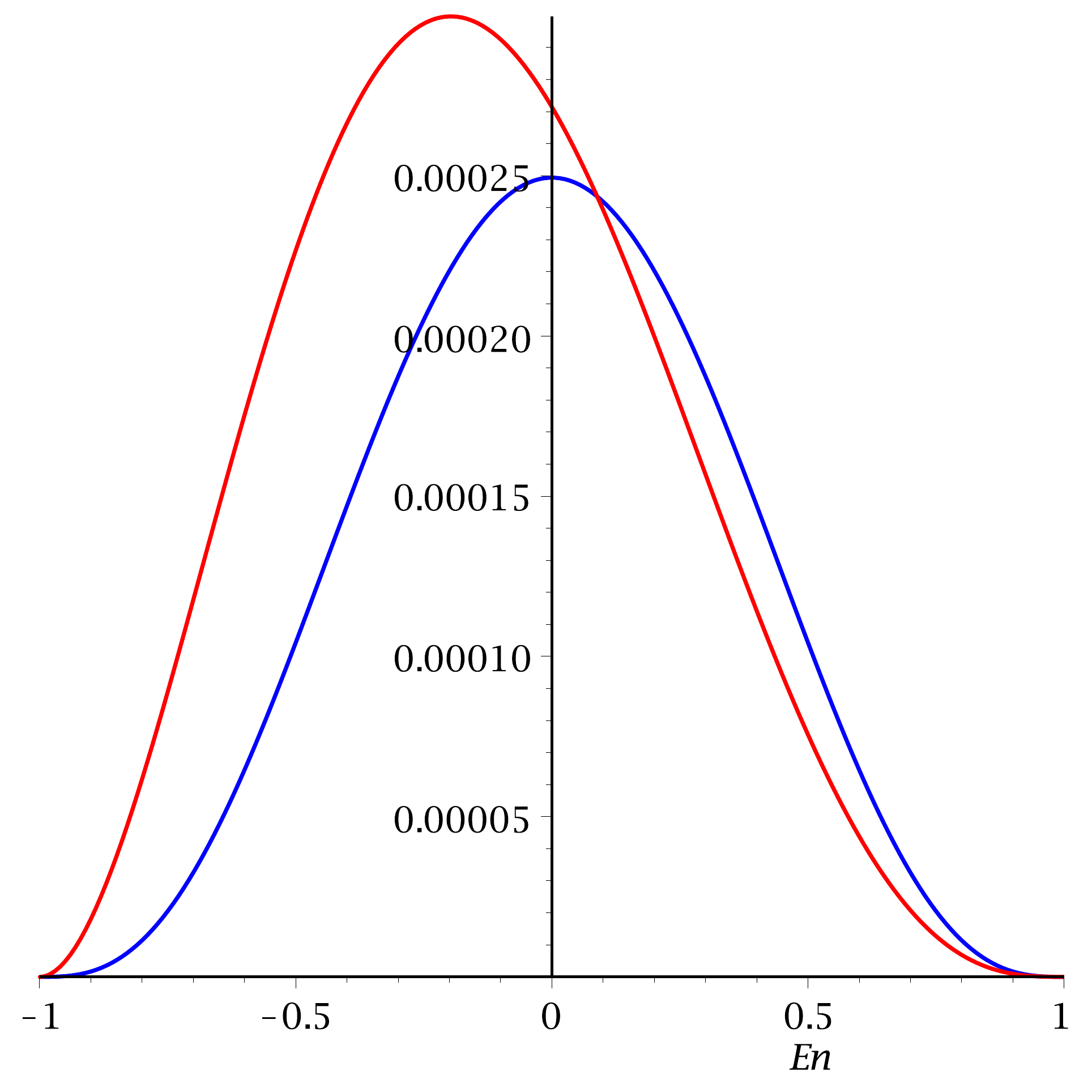} \quad  
\includegraphics[width=0.3\columnwidth]{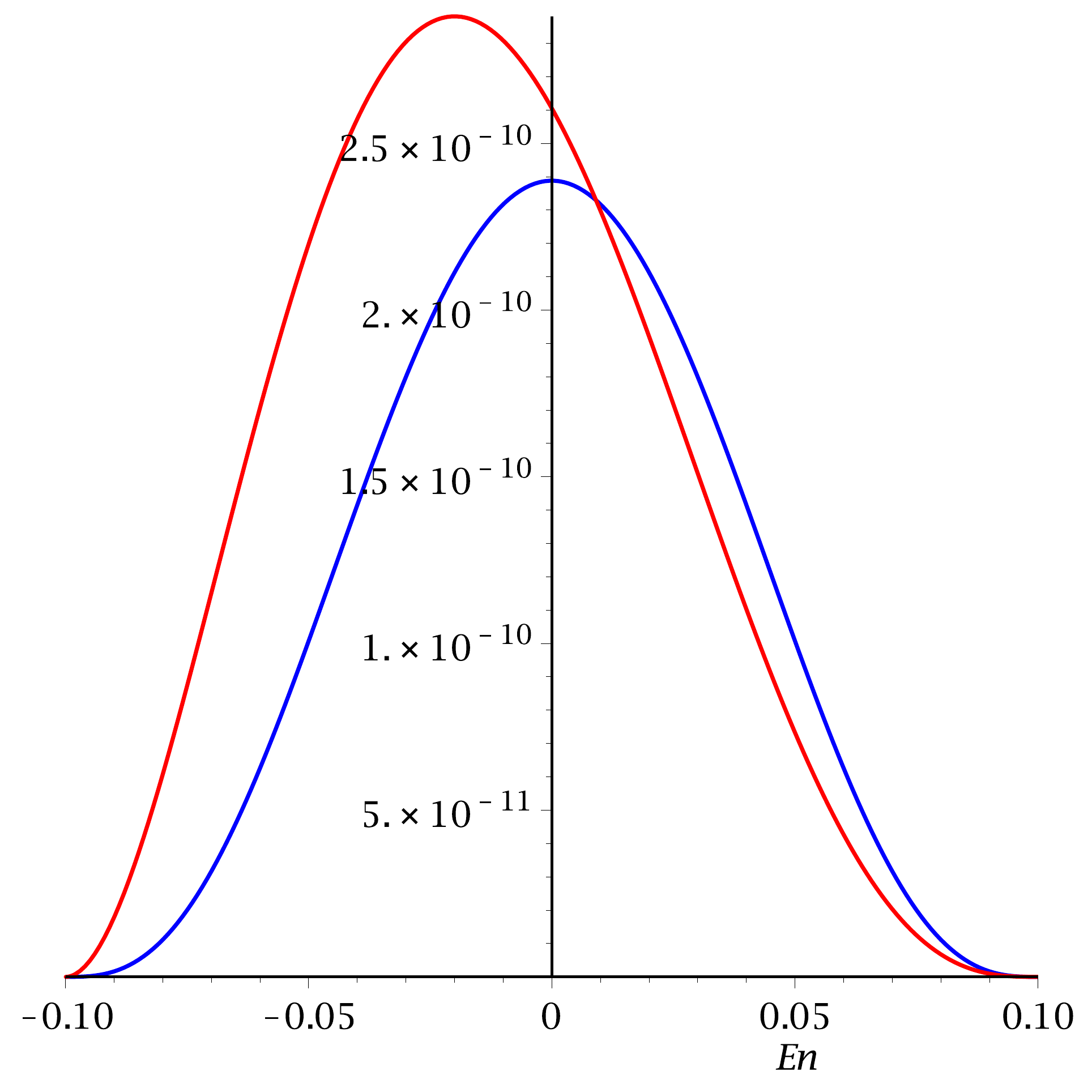} \quad  
\includegraphics[width=0.3\columnwidth]{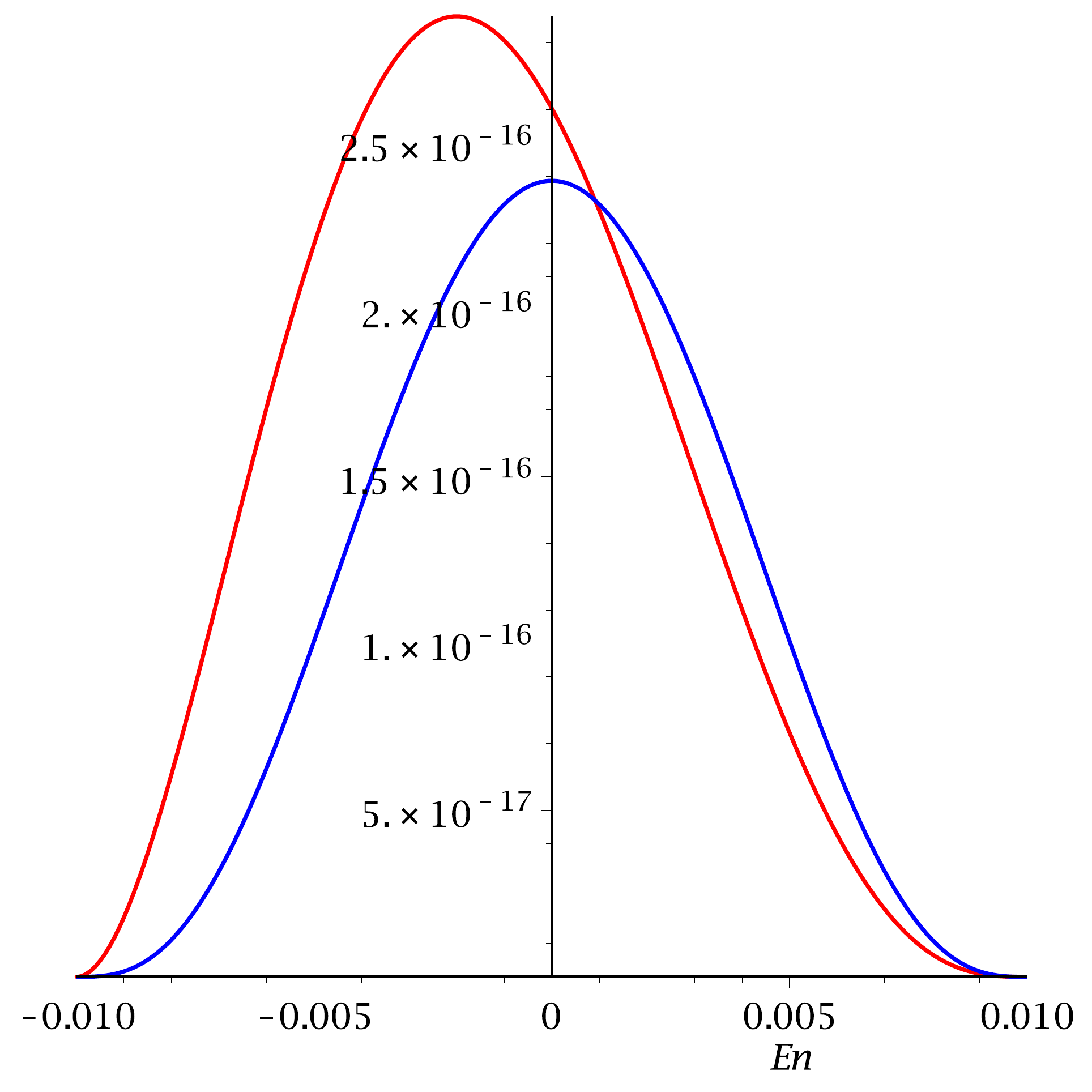}  
\end{center}  
\caption{Graphical solution of equition (\protect{\ref{eigen-energy eq(t<0)}}) for $j=11/2$, $R=1$  
and for $t=-1, -0.1, -0.01$.}  
\label{Fig j>0,t<0, BesselI}   
\end{figure}

In the rest of this section, we deal with the boundary condition   
$\Phi_j(R,\theta)\in \mathcal{H}^{(+)}(\partial D^2_R)$.   
For this boundary condition with $t>0$, we have, in place of \eqref{bdy-cond (-)},     
\begin{equation}   
\label{bdy-cond (+)}  
     \begin{pmatrix}   
     \sqrt{t+E_j}e^{i(j-\frac12)\theta}I_{j-\frac12}(\varepsilon_j R) \\  
                                               -i\sqrt{t-E_j}e^{i(j+\frac12)\theta}I_{j+\frac12}(\varepsilon_j R)   
     \end{pmatrix} = c \begin{pmatrix} -it\, e^{i(j-\frac12)\theta} \\   
                               (\frac{j}{R}+\lambda_j^{+}) e^{i(j+\frac12)\theta} \end{pmatrix} ,  
\end{equation}  
which gives rise, instead of \eqref{eigen-energy eq}, to   
\begin{equation}  
\label{eigen-energy eq(-)}   
    t\sqrt{\frac{t+E_j}{t-E_j}} I_{j-\frac12}(\varepsilon_j R) =   
    \Bigl(\frac{j}{R}-\sqrt{\frac{j^2}{R^2}+t^2}\Bigr) I_{j+\frac12}(\varepsilon_j R).   
\end{equation}   
Since the factor $\frac{j}{R}-\sqrt{\frac{j^2}{R^2}+t^2}$ in the right-hand side of the above equation   
is negative for any $j$, positive or negative, and since the other factors are all positive,   
the above equation has no solution.   
In the case of $t<0$, the boundary condition in question provides   
\begin{equation}  
\label{eigen-energy eq(-)(t<0)}   
    |t|\sqrt{\frac{t+E_j}{t-E_j}} I_{j-\frac12}(\varepsilon_j R) =   
    \Bigl(\frac{j}{R}-\sqrt{\frac{j^2}{R^2}+t^2}\Bigr) I_{j+\frac12}(\varepsilon_j R),  
\end{equation}     
which corresponds to \eqref{eigen-energy eq(t<0)}.   
This equation has no solution, either.   
We have thus found that the boundary condition $\Phi_j(R,\theta)\in \mathcal{H}^{(+)}(\partial R^2_R)$   
yields no non-trivial solution.   
\section{Zero modes}  
\label{zero modes}  
\par  
We are here interested in zero mode solutions associated with the zero eigenvalue of the Hamiltonian $\hat{H}_t$.   
Though we have obtained the modified Bessel equations \eqref{mod Bessel eqs} with the condition    
$|E_j|<|t|$,  in order to obtain  differential equations in the limit as $t\to 0$ within the constraint $|E_j|<|t|$,   
we have to take the condition $E_j=t=0$ into account.     
From \eqref{Eq-phis} along with $E_j=t=0$, we obtain   
\begin{subequations}  
\label{Eq-phis(0)}  
\begin{align}  
    -i\frac{d\phi^{(+)}_j}{dr}-\frac{i}{r}(j+\frac12)\phi^{(+)}_j=0, \label{Eq-phi+(0)}\\  
    -i\frac{d\phi^{(-)}_j}{dr}+\frac{i}{r}(j-\frac12)\phi^{(-)}_j =0. \label{Eq-phi-(0)}  
\end{align}  
\end{subequations}  
These equations are easily solved to give   
\begin{equation}   
             \phi_j^{(-)}(r)=C_1 r^{j-\frac12}, \quad \phi_j^{(+)}(r)=C_2 r^{-(j+\frac12)}.   
\end{equation}   
According as  $j>0$ or $j<0$, one should take $C_2=0$ or $C_1=0$   
because of the boundedness of $\phi^{(\pm)}_j$ as $r\to 0$.    
Thus, we find that solutions to $\hat{H}_0\Phi_j=0$ should take the form   
\begin{subequations}  
\label{zero-modes}  
 \begin{align}   
    \Phi_j(r,\theta) & =c\begin{pmatrix}   
          0 \\ e^{i(j+\frac12)\theta}r^{-(j+\frac12)} \end{pmatrix} \quad {\rm for} \quad j<0, \\  
   \Phi_j(r,\theta) & = c' \begin{pmatrix}   
          e^{i(j-\frac12)\theta}r^{j-\frac12} \\ 0 \end{pmatrix} \quad {\rm for} \quad j>0.   
 \end{align}  
\end{subequations}  
\medskip\par\noindent  
[Remark]  To see the meaning of zero modes, we introduce a complex variable $z=q_1+iq_2$   
and rewrite these solutions in terms of $z, \bar{z}$.    
The Hamiltonian $\hat{H}_0$ is rewritten as   
\begin{equation}   
    \hat{H}_0=\begin{pmatrix} 0 & -i\frac{\partial}{\partial q_1}-\frac{\partial}{\partial q_2} \\  
          -i\frac{\partial}{\partial q_1}+\frac{\partial}{\partial q_2} & 0 \end{pmatrix} =   
     \begin{pmatrix} 0 & -2i \frac{\partial}{\partial z} \\ -2i \frac{\partial}{\partial \bar{z}} & 0 \end{pmatrix} ,  
\end{equation}   
and the functions given in \eqref{zero-modes} are put in the form   
\begin{subequations}  
\label{zero-modes(2)}  
 \begin{align}   
    \Phi_j(r,\theta) & =c\begin{pmatrix}   
          0 \\ e^{i(j+\frac12)\theta}r^{-(j+\frac12)} \end{pmatrix} =   
           c \begin{pmatrix} 0 \\ \bar{z}^{|j|-\frac12} \end{pmatrix}   
     \quad {\rm for} \quad j<0, \\  
   \Phi_j(r,\theta) & = c' \begin{pmatrix}   
          e^{i(j-\frac12)\theta}r^{j-\frac12} \\ 0 \end{pmatrix}   
      = c'\begin{pmatrix} z^{j-\frac12} \\ 0 \end{pmatrix}   
\quad {\rm for} \quad j>0.   
 \end{align}  
\end{subequations}  
The component functions,  $\bar{z}^{|j|-\frac12},  z^{j-\frac12}$,    
are anti-holomorphic and holomorphic functions of special symmetry (where    
it is to be noted that $|j|-\frac12$ and $j-\frac12$ are integers),   
which are kernels of $\partial/\partial z$ and $\partial/\partial \bar{z}$, respectively,   
and solutions of the Laplace equation $\Delta u=4\frac{\partial^2}{\partial z \partial \bar{z}}u=0$ as well.   
  
\medskip\par  
To obtain actual solutions, we have to take suitable boundary conditions into account.   
To this end, we study the decomposition of $\mathcal{H}(\partial D^2_R)$   
with respect to the operator $K_0$ with $t=0$.    
Since $K_0$ is a diagonal operator, it is easy to find the eigenvalues and the associated eigenstates,     
\begin{subequations}  
\label{bdy zero modes(j)}  
 \begin{align}  
    K_0\phi^{(0,+)}_j & =\frac{1}{R}(j+\frac12)\phi^{(0,+)}_{j}, \quad   
       \phi^{(0,+)}_j=\begin{pmatrix} 0 \\ a_j e^{i(j+\frac12)\theta} \end{pmatrix}, \\   
    K_0\phi^{(0,-)}_j & =-\frac{1}{R}(j-\frac12)\phi^{(0,-)}_j, \quad   
       \phi^{(0,-)}_j=\begin{pmatrix}  b_j e^{i(j-\frac12)\theta} \\ 0 \end{pmatrix}.   
 \end{align}  
\end{subequations}  
These eigenvalues are obtained also from \eqref{lambda(pm)_j} with $t=0$;   
\begin{equation}  
\label{K0-eigenvalues}   
    \kappa^{\pm}_j|_{t=0}=\frac{1}{2R}+\lambda^{\pm}_j\Bigr|_{t=0}=\frac{1}{R}(\frac12\pm |j|) . 
\end{equation}  
Let $\mathcal{H}^{(+)}_0(\partial D^2_R)$ and $\mathcal{H}^{(-)}_0(\partial D^2_R)$ be   
spaces  spanned by   
$\phi^{(0,+)}_j;\, j=\pm\frac12, \pm\frac32,\cdots$ and by    
$\phi^{(0,-)}_j;  j=\pm\frac12, \pm\frac32,\cdots$, respectively.   
Then, the  $\mathcal{H}(\partial D^2_R)$ is decomposed into the direct sum        
$\mathcal{H}(\partial D^2_R)=\mathcal{H}^{(+)}_0(\partial D^2_R)\oplus \mathcal{H}^{(-)}_0(\partial D^2_R)$   
with respect to $K_0$.   
Though the superscripts $(\pm)$ are used to distinguish two subspaces, they do not indicate that the   
eigenvalues concerned are positive or negative.   
Both $\mathcal{H}^{(\pm)}_0(\partial D^2_R)$ have eigenstates associated with   
negative, zero, and positive eigenvalues of $K_0$, as is seen from \eqref{bdy zero modes(j)}.    
For $j=-\frac12$ and $j=\frac12$, the subspaces $\mathcal{H}^{(+)}_0(\partial D^2_R)$ and   
$\mathcal{H}^{(-)}_0(\partial D^2_R)$ contain the eigenstate associated with zero eigenvalue,  respectively.       
    
As for the action of $\sigma_r$ on $\mathcal{H}^{(\pm)}_0(\partial D^2_R)$,   
a straightforward calculation shows that $\sigma_r \phi^{(0,-)}_j={\rm const} \phi^{(0,+)}_j$ and   
$\sigma_r \phi^{(0,+)}_j={\rm const} \phi^{(0,-)}_j$.   
This implies that   
\begin{equation}   
  \sigma_r \mathcal{H}^{(\pm)}_0(\partial D^2_R)=\mathcal{H}^{(\mp)}_0(\partial D^2_R).  
\end{equation}   
It is to be noted that $\mathcal{H}^{(+)}_0(\partial D^2_R)$ and $\mathcal{H}^{(-)}_0(\partial D^2_R)$ are   
orthogonal.   
This fact ensures that the Hamiltonian $\hat{H}_0$ is symmetric if spinor states on $D^2_R$ are subject to   
the boundary condition that $\Phi(R,\theta)$ belongs to either of $\mathcal{H}^{(\pm)}_0(\partial D^2_R)$.   
  
We have two choices of boundary conditions for $\Phi(R,\theta)$, which are   
$\Phi(R,\theta)\in \mathcal{H}^{(+)}_0(\partial D^2_R)$ and   
$\Phi(R,\theta)\in \mathcal{H}^{(-)}_0(\partial D^2_R)$.    
In view of \eqref{bdy zero modes(j)} and \eqref{zero-modes}, we have to choose   
a suitable one depending on the sign of $j$.   
For example, if we choose the boundary condition $\Phi_j(R,\theta)\in \mathcal{H}^{(-)}_0(\partial D^2_R)$   
for $j<0$, we have trivial solutions.  In contrast with this, the boundary condition   
$\Phi_j(R,\theta)\in \mathcal{H}^{(+)}_0(\partial D^2_R)$ gives a non-trivial solution.   
In the case of $j>0$, we have to choose the boundary condition   
$\Phi_j(R,\theta)\in \mathcal{H}^{(-)}_0(\partial D^2_R)$ for a non-trivial solution.    
  
\section{Regular states}   
\label{regular states}  
\par  
In this section, we apply the APS boundary condition \eqref{APS Disk} to feasible solutions   
which are described in terms of Bessel functions in the case of $|E_j|>|t|$.    
Recall that those solutions are given by \eqref{sol |E|>|t|}, depending on $E_j>0$ and $E_j<0$.   
  
(I) First we adopt the APS boundary condition given by   
$\Phi_j(R, \theta)\in \mathcal{H}^{(-)}(\partial D^2_R)$.   
For $E_j>0$, the boundary condition is expressed as   
\begin{equation}   
        \begin{pmatrix} \sqrt{E_j+t}e^{i(j-\frac12)\theta}J_{j-\frac12}(\beta_j R) \\  
                                               i\sqrt{E_j-t}e^{i(j+\frac12)\theta}J_{j+\frac12}(\beta_j R)  \end{pmatrix}  
  = c \begin{pmatrix}   
    -it e^{i(j-\frac12)\theta}  \\ (\frac{j}{R}+\lambda^-_j)e^{i(j+\frac12)\theta} \end{pmatrix} .   
\end{equation}   
This equation yields   
\begin{equation}  
\label{E>0, Bessel(-)}   
  -t\sqrt{\frac{E_j+t}{E_j-t}}J_{j-\frac12}(\beta_j R) =   
  \Bigl(\frac{j}{R}+\sqrt{\frac{j^2}{R^2}+t^2}\Bigr) J_{j+\frac12}(\beta_j R).   
\end{equation}   
For $E_j<0$, we obtain the boundary condition   
\begin{equation}   
        \begin{pmatrix} \sqrt{|E_j+t|}e^{i(j-\frac12)\theta}J_{j-\frac12}(\beta_j R) \\  
                                               -i\sqrt{|E_j-t|}e^{i(j+\frac12)\theta}J_{j+\frac12}(\beta_j R)  \end{pmatrix}  
  = c \begin{pmatrix}   
    -it e^{i(j-\frac12)\theta}  \\ (\frac{j}{R}+\lambda^-_j)e^{i(j+\frac12)\theta} \end{pmatrix} ,   
\end{equation}   
which is arranged to give   
\begin{equation}  
\label{E<0, Bessel(-)}  
   t\sqrt{\frac{|E_j+t|}{|E_j-t|}}J_{j-\frac12}(\beta_j R) =   
  \Bigl(\frac{j}{R}+\sqrt{\frac{j^2}{R^2}+t^2}\Bigr) J_{j+\frac12}(\beta_j R).   
\end{equation}   
  
(II) We turn to the other APS boundary $\Phi_j(R, \theta)\in \mathcal{H}^{(+)}(\partial D^2_R)$.  
For $E_j>0$, the APS boundary condition is written as   
\begin{equation}   
        \begin{pmatrix} \sqrt{E_j+t}e^{i(j-\frac12)\theta}J_{j-\frac12}(\beta_j R) \\  
                                               i\sqrt{E_j-t}e^{i(j+\frac12)\theta}J_{j+\frac12}(\beta_j R)  \end{pmatrix}  
  = c \begin{pmatrix}   
    -it e^{i(j-\frac12)\theta}  \\ (\frac{j}{R}+\lambda^+_j)e^{i(j+\frac12)\theta} \end{pmatrix} .   
\end{equation}   
This gives rise to   
\begin{equation}  
\label{E>0, Bessel(+)}  
   -t\sqrt{\frac{|E_j+t|}{|E_j-t|}}J_{j-\frac12}(\beta_j R) =   
  \Bigl(\frac{j}{R}-\sqrt{\frac{j^2}{R^2}+t^2}\Bigr) J_{j+\frac12}(\beta_j R).   
\end{equation}   
For $E_j<0$, the present APS boundary condition is expressed as   
\begin{equation}   
        \begin{pmatrix} \sqrt{|E_j+t|}e^{i(j-\frac12)\theta}J_{j-\frac12}(\beta_j R) \\  
                                               -i\sqrt{|E_j-t|}e^{i(j+\frac12)\theta}J_{j+\frac12}(\beta_j R)  \end{pmatrix}  
  = c \begin{pmatrix}   
    -it e^{i(j-\frac12)\theta}  \\ (\frac{j}{R}+\lambda^+_j)e^{i(j+\frac12)\theta} \end{pmatrix} ,   
\end{equation}    
which is arranged to give   
\begin{equation}  
\label{E<0, Bessel(+)}   
   t\sqrt{\frac{|E_j+t|}{|E_j-t|}}J_{j-\frac12}(\beta_j R) =   
  \Bigl(\frac{j}{R}-\sqrt{\frac{j^2}{R^2}+t^2}\Bigr) J_{j+\frac12}(\beta_j R).   
\end{equation}   
  
The four defining equations, \eqref{E>0, Bessel(-)},  \eqref{E<0, Bessel(-)}, \eqref{E>0, Bessel(+)},   
\eqref{E<0, Bessel(+)}, for eigenvalues have solutions, which can be graphically verified, as was done   
for edge states.

\section{Spectral flow for the local full quantum system}  
\label{spectral flow}  
\par  
In order to observe the transition of the eigenvalues against $t$ for edge states,   
we have to choose a suitable boundary condition out of $\mathcal{H}^{(\pm)}_0(\partial D^2_R)$   
for zero modes.  To this end,   
we sum up the discussion on the possibility of energy eigenvalues for edge states (Sec.~\ref{edge states})   
to obtain the following tables,   
\begin{equation}  
   \begin{array}{c|c|c|c} t\leq 0  & E_j<0 & E_j=0 & E_j>0 \\ \hline  
                              j<0 &  &  &  {\rm no} \\ \hline  
                              j>0 &  {\rm no} &  & \end{array} \qquad   
   \begin{array}{c|c|c|c}  t\geq 0 & E_j<0 & E_j=0 & E_j>0 \\ \hline  
                              j<0 &  {\rm no} &  &   \\ \hline  
                              j>0 &   &  &  {\rm no} \end{array}  
\end{equation}  
From these tables, we can find possible transitions of eigenstates against $t$ increasing from   
negative values to positive values.   
Suppose we have solutions for $t<0$.  Then, the possible pairs of $j$ and $E_j$ for the existence of   
eigenstates are (i) $j<0$ and $E_j<0$ and (ii) $j>0$ and $E_j>0$.    
On the contrary, if we have solutions for $t>0$, the possible pairs of $j$ and $E_j$ are (iii) $j>0$ and $E_j<0$,   
and (iv) $j<0$ and $E_j>0$.   
  
In the case of (i),  in the limit as $t\to 0$, we have to choose the boundary condition   
$\Phi_j(R,\theta)\in \mathcal{H}^{(+)}_0(\partial D^2_R)$   
in order to have a non-trivial zero mode solution.     
If $t$ goes upward from the $t<0$ side to the $t>0$ side,   
a possible eigenstate should be associated with a positive eigenvalue $E_j$,   
since the case (iv) must occur because of the conservation of the spin-orbital angular momentum.   
In contrast with this, in the case of (ii), we have to choose the boundary condition   
$\Phi_j(R,\theta)\in \mathcal{H}^{(-)}_0(\partial D^2_R)$ in the limit as $t\to 0$.   
When $t$ goes upward from the negative to the positive sides,    
eigenstates associated with negative energy eigenvalues should come out,   
in view of the case (iii).    
Thus, the transition of eigenstates against $t$ for edge states with transient zero modes   
is summed up in the table   
\begin{equation}  
\label{edge state transition}   
 \begin{array}{c|c|c|c}  
         & t<0 & t=0 & t>0 \\ \hline   
    j>0 & E_j>0 & E_j=0 & E_j<0 \\ \hline   
    j<0 & E_j<0 & E_j =0 & E_j >0   
 \end{array}   
\end{equation}   
This table can be considered as describing the full quantum local delta-Chern,   
which is comparable with the table (\ref{ex pt shift}) for the semi-quantum model.    
  
\begin{figure}[htbp]  
\begin{center}  
\includegraphics[width=1\columnwidth]{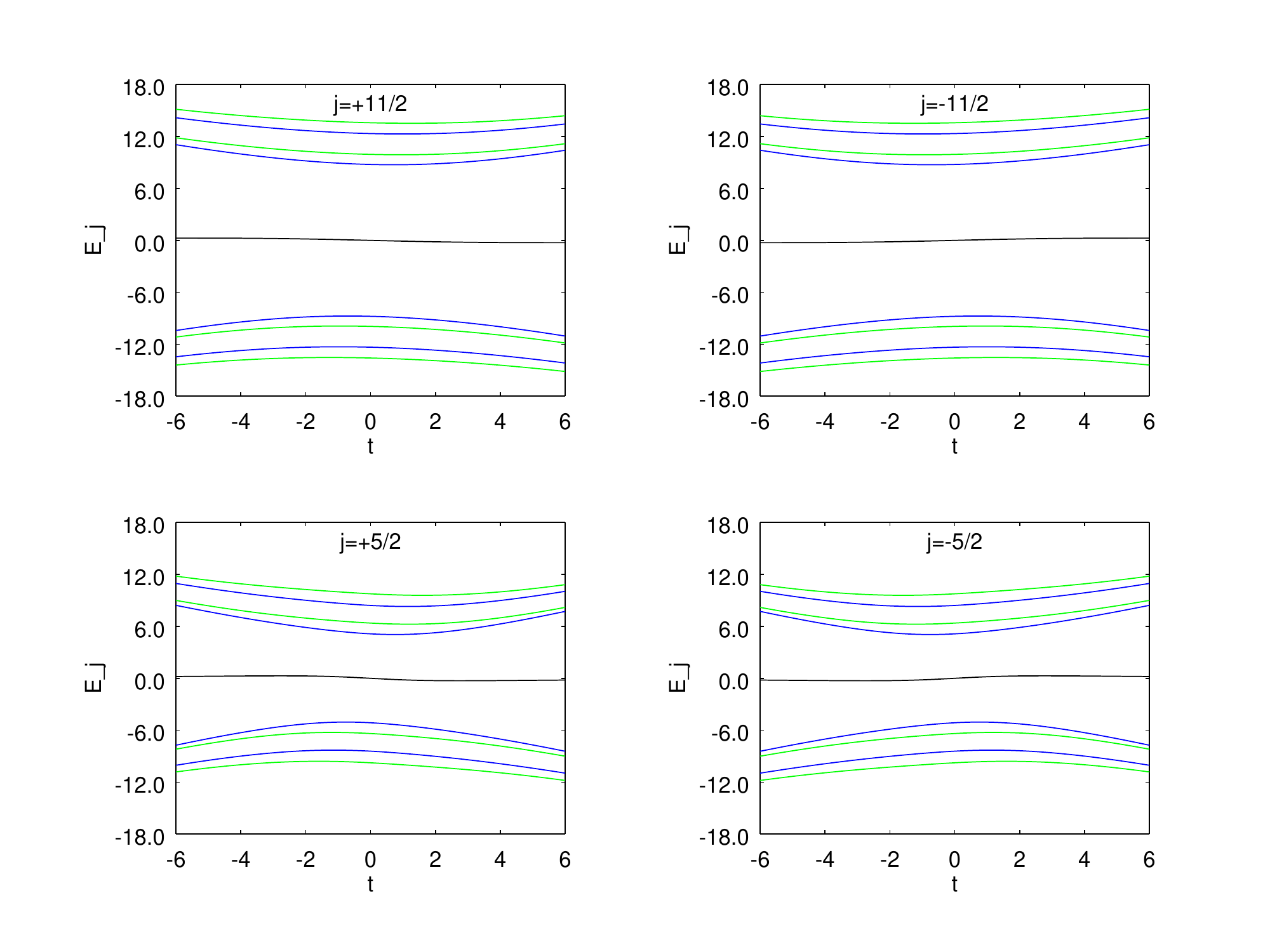}  
\end{center}  
\caption{Energy eigenvalues as functions of the parameter $t$.  Green lines are solutions  
for APS boundary conditions with $\Phi_j(R,\theta)\in {\cal H}^{(-)} (\partial D_R^2)$.   
(Eqs.~\eqref{E>0, Bessel(-)} and   
\eqref{E<0, Bessel(-)}).  
Blue lines are solutions for APS boundary conditions with $\Phi_j(R,\theta)\in {\cal H}^{(+)} (\partial D_R^2)$.   
 (Eqs.~\eqref{E>0, Bessel(+)} and \eqref{E<0, Bessel(+)}).  Black lines are solutions for edge states.  
\label{E(t)} }  
\end{figure}

In order to see the evolution of the whole patterns of eigenvalues,   
we need to observe that of eigenvalues for regular states as well.    
In contrast with the case of $|E_j|<|t|$, in the region $|E_j|>|t|$, one can make $t$ tend to zero   
irrespective of $E_j$.  This  implies that if $t$ tends to zero the Bessel differential equations   
\eqref{Bessel(pm)} remains to be so. Therefore, the defining equations of eigenvalues,    
\eqref{E>0, Bessel(-)},  \eqref{E<0, Bessel(-)}, \eqref{E>0, Bessel(+)}, \eqref{E<0, Bessel(+)},    
for regular states are valid for all parameter values near $t=0$.     
In fact, we can find numerical solutions to those equations as functions of $t$,   
as is shown in Fig.~\ref{E(t)} together with   
the numerical solutions of eigenvalues for edge states.

\begin{figure}  
\begin{center}  
\includegraphics[width=0.45\columnwidth]{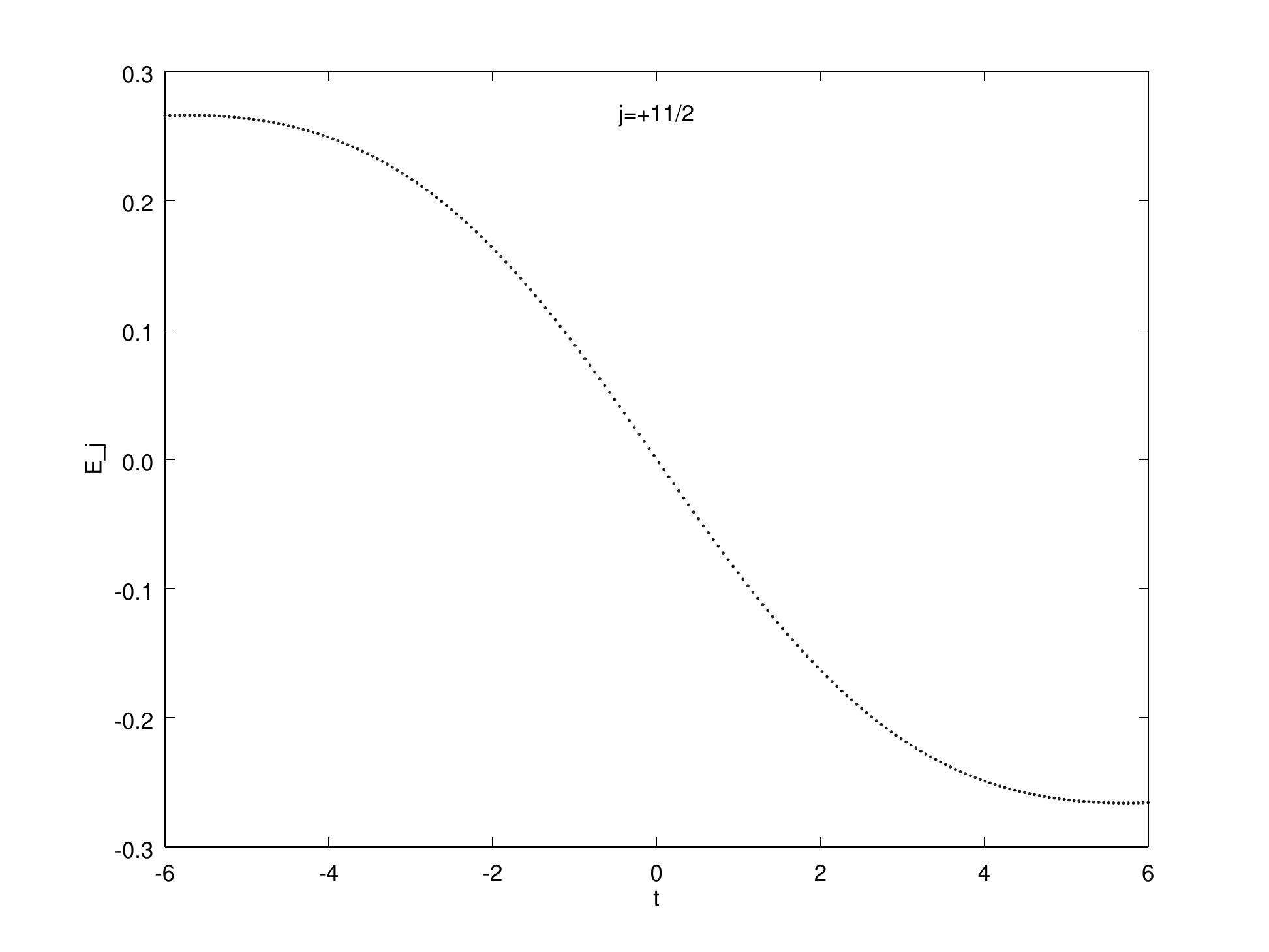} \quad   
\includegraphics[width=0.45\columnwidth]{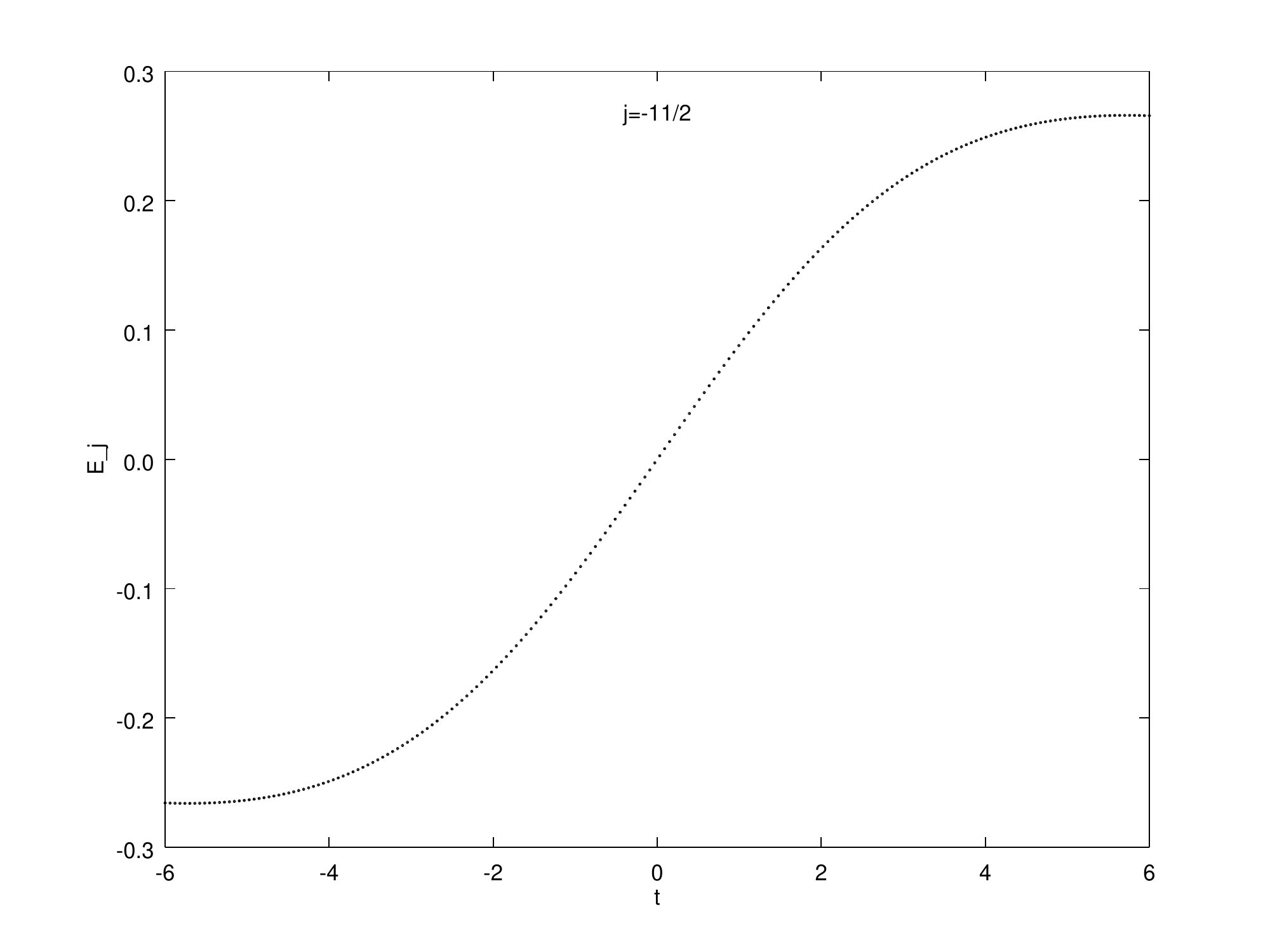} \\   
\includegraphics[width=0.45\columnwidth]{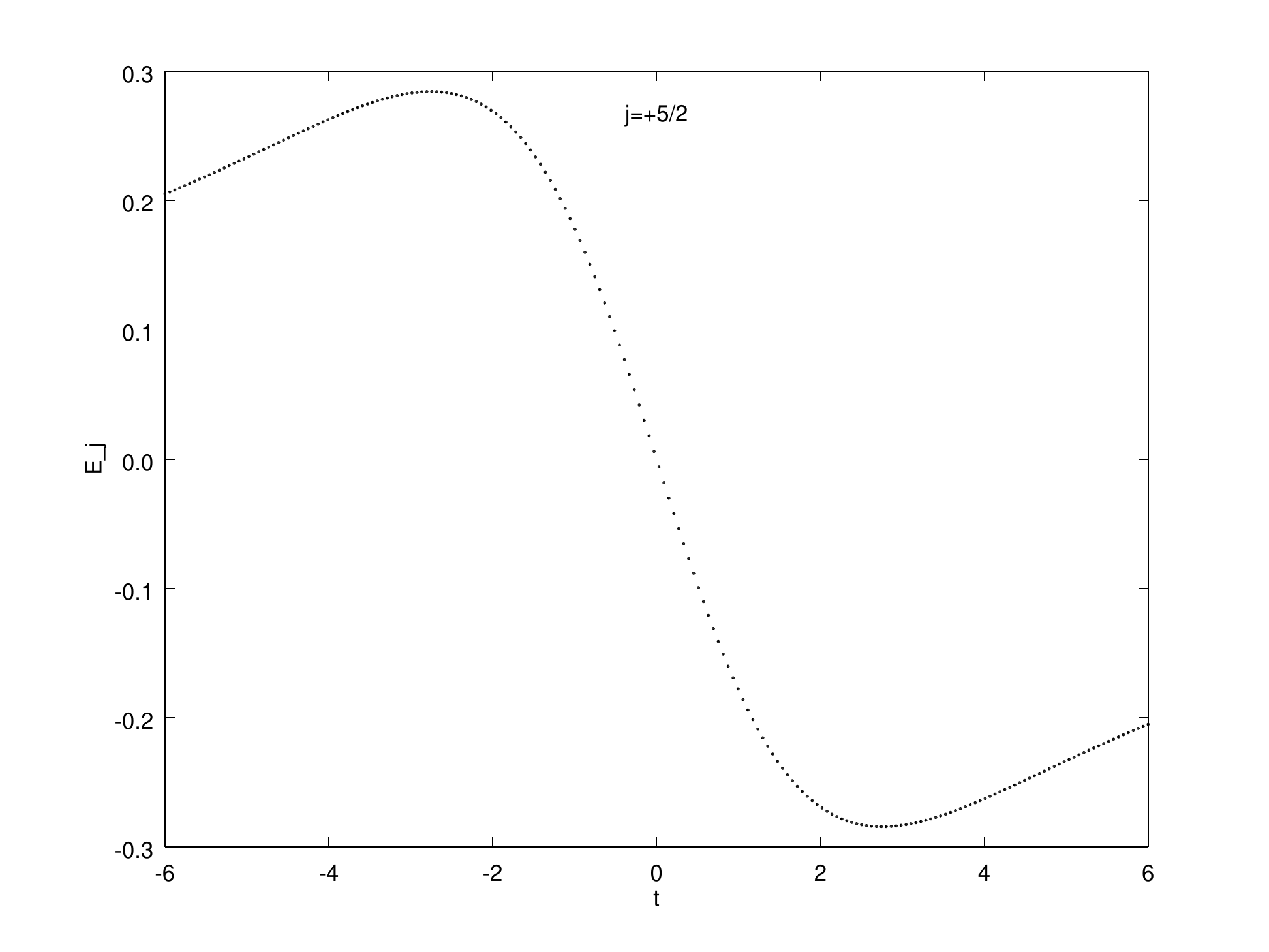} \quad   
\includegraphics[width=0.45\columnwidth]{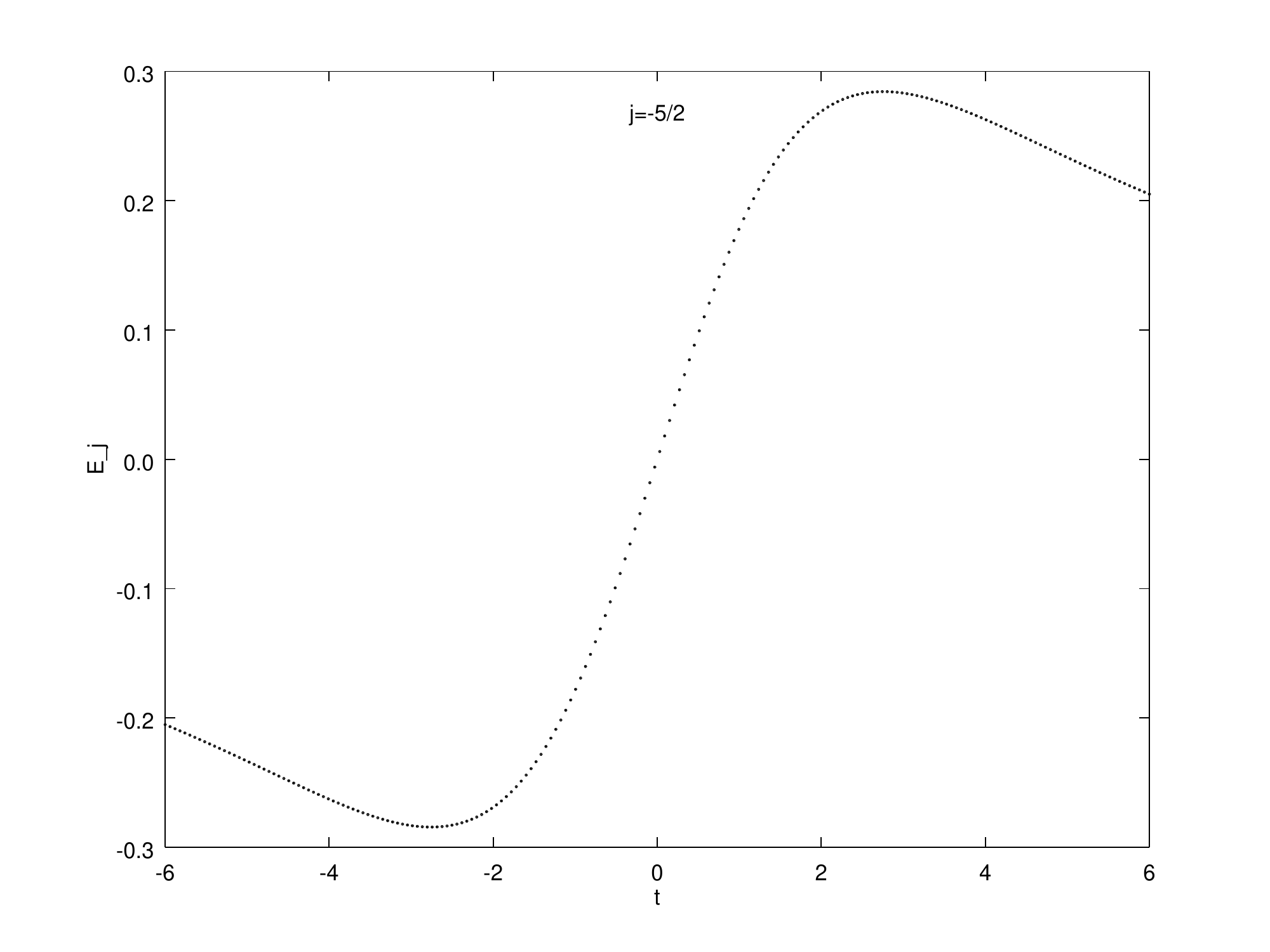}   
\end{center}  
\caption{Evolution of an edge state as a function of control parameter  
$t$ for a local Dirac model. Left graphs are for positive values of $j=11/2$ and  
$j=5/2$. Right graphs are for negatives values $j=-11/2$ and $j=-5/2$.    
\label{EofTforEdgeStates} }  
\end{figure}  
  
To have a close look at eigenvalues for edge states as functions of $t$, we add four figures   
in Fig.~\ref{EofTforEdgeStates}.   
Depending on the sign of the total angular momentum $j$,   
the edge state  goes from the lower energy band to the upper  
energy band and vice versa along with variation of $t$ parameter from  
negative to positive.  These figures are graphical realizations of  \eqref{edge state transition}.

According to \cite{APS,Prokh},  the spectral flow of a one-parameter family of self-adjoint operators   
is defined to be the net number of eigenvalues passing through zero in the positive direction as the parameter   
$t$ runs, that is, the difference between the number of eigenvalues (counting multiplicities)   
crossing zero in positive and negative directions.    
Figures \ref{E(t)}, \ref{EofTforEdgeStates}
show that the spectral flow of $\hat{H}_t$ is $\mp 1$, depending on ${\rm sgn}(j)$.

\section{Concluding remarks on global and local quantum and semi-quantum models and   
on an analogy to topological insulators}  
\label{conclusion}  
We now return back to a full quantum rotational problem for two quantum states  
in order to associate it with the characteristic features of solutions found for the local Dirac model.  
The zero mode can correspond to a quantum state which undergoes redistribution between two energy   
bands for an initial rotation-vibration problem.   
This quantum state has an extremal projection of the orbital angular momentum,   
$\pm J$,  on the axis going through the degeneracy point of the semi-quantum model.   
This means that on the sphere  $S^2$ the corresponding eigenstate is localised   
along a great circle situated maximally far from the axis going through the degeneracy point.   
  
The redistributing  state (edge state) is assigned to one (upper in energy) or   
to another (lower in energy) band, depending on positive or negative value of the control parameter $t$.   
From the viewpoint of the local Dirac model, the state can be assigned to one or to another band,   
depending on whether the average value of $\sigma_\theta$ is positive or negative.     
We can check these properties for the boundary states $\mathcal{H}^{(-)}(\partial D^2_R)$.   
As is already known, the radial component of the spin vanishes;   
$\langle \phi^{(-)}_j, \sigma_r \phi^{(-)}_j\rangle_S =0$.   
We now evaluate the tangential component of the spin.    
A straightforward calculation gives    
\begin{equation}  
   \langle \phi^{(-)}_j, \sigma_{\theta}\phi^{(-)}_j \rangle_S =  
    2\pi |c'_j|^2t\Bigl(\frac{j}{R}-\sqrt{\frac{j^2}{R^2}+t^2}\Bigr)   
  \;\left\{\begin{array}{ccc} < 0 & {\rm for} & t>0, \\   
                               > 0 & {\rm for} & t<0,  \end{array} \right.   
\end{equation}  
independently of $j$. This confirms that the boundary state goes from one  
band to another along with parameter $t$ going through zero.  
For the comparison's sake, we calculate the average of the orbital momentum to find that   
\begin{equation}   
   \langle \phi^{(-)}_j, (-i\partial_{\theta})\phi^{(-)}_j \rangle_S  \;  
   \left\{ \begin{array}{ccc} >0 & {\rm for} & j>0, \\  
                                    <0 & {\rm for} & j<0, \end{array} \right.   
\end{equation}  
independently of $t$.   
  
In conclusion, we add further graphs to illustrate how the band rearrangement theory   
and the topological insulator theory are related together.    
They can share the same Dirac model.

\begin{figure}[htbp]  
\begin{center}  
\includegraphics[width=0.28\columnwidth]{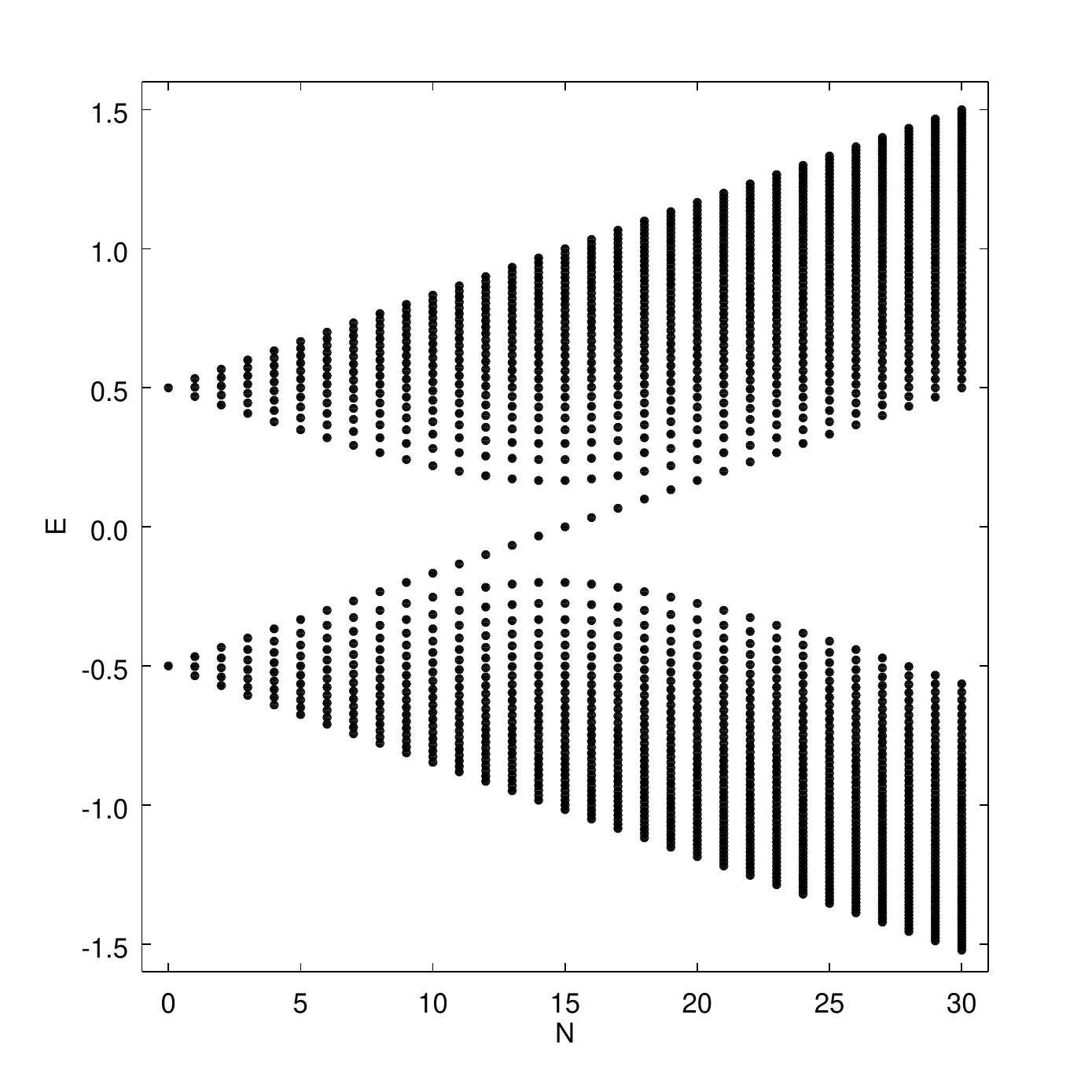} \quad   
\includegraphics[width=0.34\columnwidth]{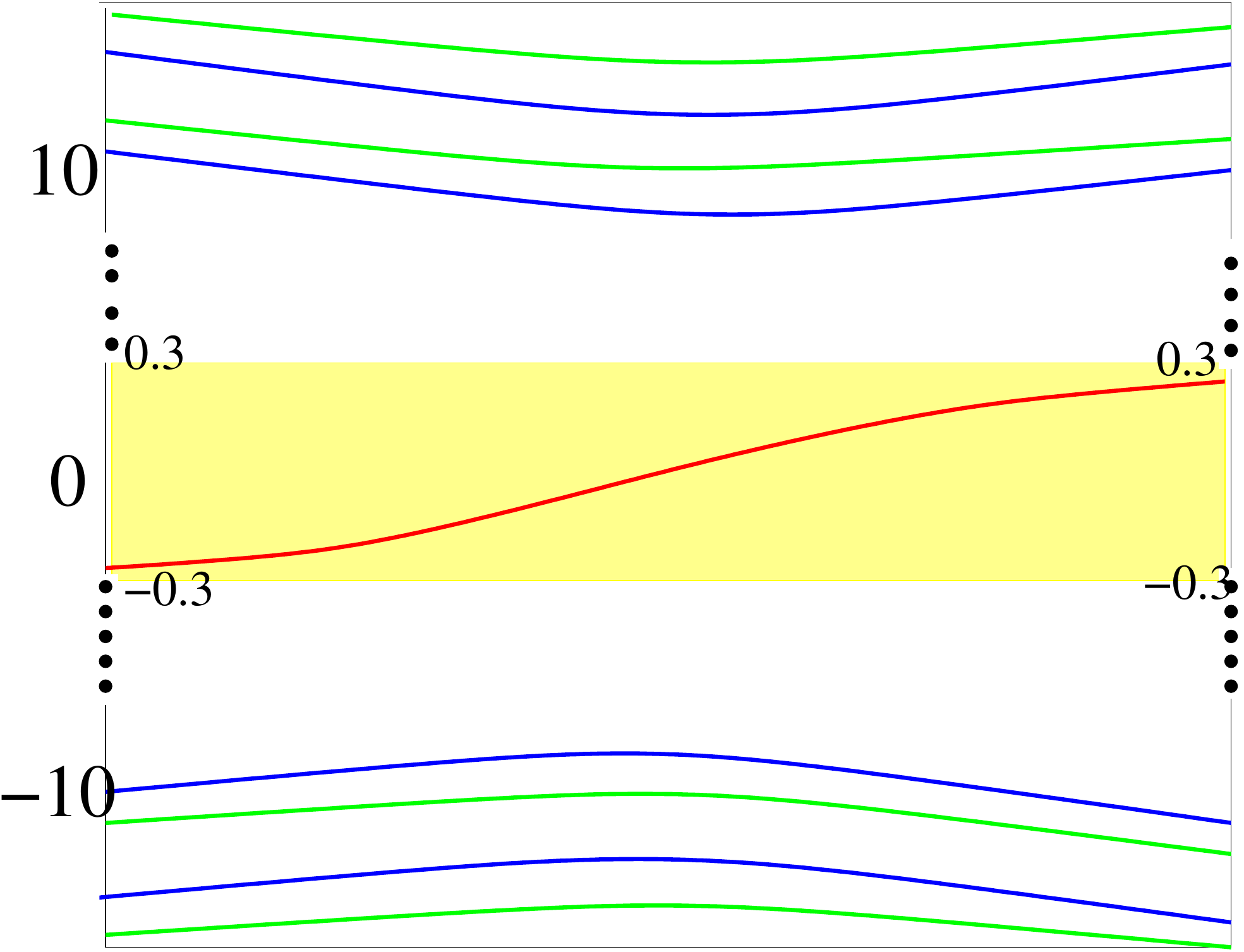} \quad  
\includegraphics[width=0.32\columnwidth]{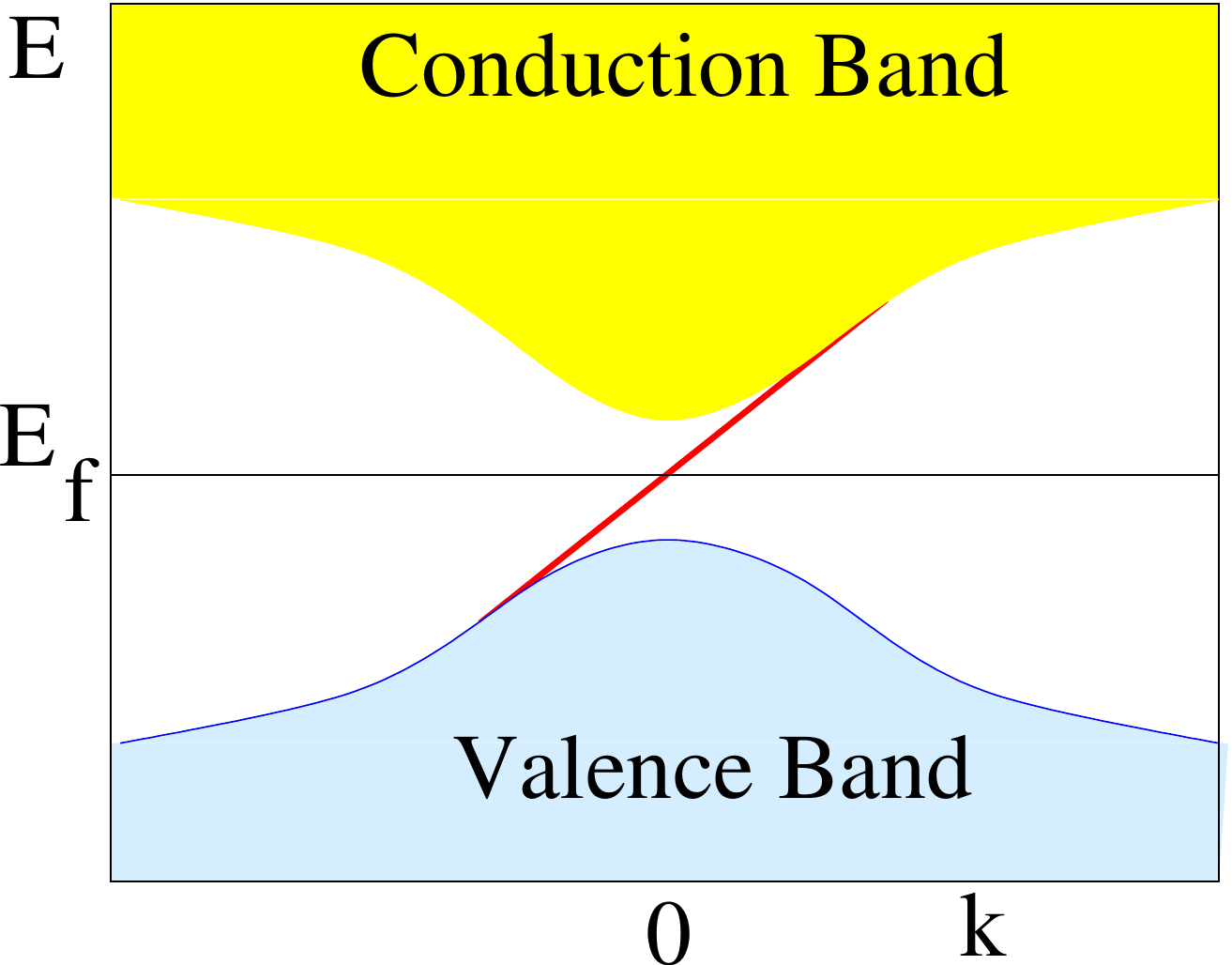}   
\end{center}  
\caption{Comparison of the redistribution of quantum energy levels between  
energy bands in isolated molecules (left subfigure), electronic states in  
two-dimensional solid as functions on the Brillouin cell (right subfigure), and the behavior  
of edge states for the local Dirac model studied in the present paper  
as functions of a control parameter $t$ (center subfigure), where different energy scales  
are used to show noticeably both regular and edge states in the center subfigure.   
\label{ComparFig} }  
\end{figure}  
  
The leftmost subfigure of Fig.~\ref{ComparFig} shows   
the phenomenon of the redistribution of energy levels between energy bands,    
which  was initially introduced in \cite{VPVdp}.    
The evolution of the band structure for two rotational bands is described as functions of   
a control parameter $N$ which is taken as the quantum number associated with   
the rotational angular momentum. In a more formal mathematical way, $N$ labels the irreducible  
representation of the $SO(3)$ group for a quantum problem but for  
semi-quantum description $N$ characterizes variables treated  
as classical ones along with the associated reduced classical phase space being   
a two-dimensional sphere.   
The model quantum Hamiltonian is taken to be  
\begin{eqnarray}  
H_{\rm model} = S_z\otimes \1 + \alpha \mathbf{S}\otimes\mathbf{N}  
\end{eqnarray}  
with $\alpha=\frac{1}{15}$. The zero energy solution corresponds in this  
model to $N=15$ and the projection of the rotational angular momentum  
on the $z$ axis at the degeneracy point is $N_z=-15$, where   
the degeneracy point  formed for $\alpha=\frac{1}{15}$ is situated on the $z$ axis.      
  
The rightmost figure of Fig.~\ref{ComparFig} is one of frequently cited graphs   
in the field of topological insulators.   
Here the momentum $k$ characterizing electrons moving in a periodic  
two-dimensional structure is looked upon as a continuous classical variable and the  
corresponding classical phase space is a two-dimensional torus.   
This figure shows the behavior of edge states appearing for solid state models  
\cite{Haldane,TopInsulRMP}.  
  
The center figure of Fig.~\ref{ComparFig} is a summary of the graphs shown in   
Figs.~\ref{E(t)} and \ref{EofTforEdgeStates}, which is drawn on the local Dirac model.   
From the viewpoint of this figure, the transient state of the band rearrangement depicted in   
the left figure is realized in the center figure, and the right figure is looked upon as a counterpart of   
the center figure as a realization in the topological insulator models.

\begin{acknowledgements}  
The authors would like to thank Dr. G. Dhont for drawing graphs of eigenvalues as functions of   
the control parameter.   
\end{acknowledgements}

\end{document}